\documentclass[]{myaa}
\usepackage{balance}
\usepackage{graphicx}
\usepackage[varg]{txfonts}
\usepackage{natbib}
\bibpunct{(}{)}{;}{a}{}{,} 
\usepackage[figuresright]{rotating}
\usepackage[a4paper,breaklinks]{hyperref}

\def\kms{$\mathrm{km\, s^{-1}}$}
\def\ms{$\mathrm{m\, s^{-1}}$ }

\newcommand{\Teff}{\ensuremath{T_\mathrm{eff}}}

\newcommand{\loggf}{\ensuremath{\log\,gf}}
\newcommand{\logg}{\ensuremath{\log g}}
\newcommand{\pun}[1]{\,#1}

\newcommand{\moh}{\ensuremath{[\mathrm{M/H}]}}

\newcommand{\draftflag}{false}

\newcommand{\beq}{\begin{equation}}
\newcommand{\eeq}{\end{equation}}

\newcommand{\COBOLD}{{\sf CO$^5$BOLD}}

\newcommand{\cobold}{\COBOLD}
\newcommand{\linfor}{{\sf Linfor3D}}

\newcommand{\xx}{\ensuremath{\mathrm{1D}_{\mathrm{LHD}}}}
\newcommand{\mD}{\ensuremath{\left\langle\mathrm{3D}\right\rangle}}

\idline{1}{1}
\begin{document}

\title{The photospheric solar oxygen project\thanks{This paper
is dedicated to the memory of Hartmut Holweger.}:
I. Abundance analysis of atomic lines and influence of atmospheric models}

\author{
E.~Caffau,     \inst{1}
H.-G. Ludwig,  \inst{2,1}
M. Steffen,    \inst{3}
T.R. Ayres,    \inst{4}
P. Bonifacio,  \inst{2,1,5}
R. Cayrel,     \inst{6}
B. Freytag,    \inst{7,8}
B. Plez        \inst{8,9}}
\institute{ 
GEPI, Observatoire de Paris, CNRS, Universit\'e Paris Diderot; 92195
Meudon Cedex, France
\and
CIFIST Marie Curie Excellence Team
\and
Astrophysikalisches Institut Potsdam, An der Sternwarte 16, D-14482 Potsdam, Germany
\and
Center for Astrophysics and Space Astronomy, University of Colorado
389 UCB  (CASA),  Boulder, CO 80309-0389
\and
Istituto Nazionale di Astrofisica,
Osservatorio Astronomico di Trieste,  Via Tiepolo 11,
I-34143 Trieste, Italy
\and
GEPI, Observatoire de Paris, CNRS, Universit\'e Paris Diderot; 61 av. de l'Observatoire, F-75014 
Paris,  France
\and
CRAL,UMR 5574: CNRS, Universit\'e de Lyon,
\'Ecole Normale Sup\'erieure de Lyon,
46 all\'ee d'Italie, F-69364 Lyon Cedex 7, France
\and
Department of Physics and Astronomy, Uppsala University, Box 515, S-751 
20 Uppsala, Sweden
\and
GRAAL, CNRS UMR 5024, Universit{\'e} Montpellier~II, 34095~Montpellier~Cedex~5, France
}
\authorrunning{Caffau et al.}
\titlerunning{The solar oxygen abundance}
\offprints{E.~Caffau}
\date{Received 1 April 2008 / Accepted 28 May 2008}

\abstract%
{The solar oxygen abundance has undergone a major downward revision
in the last decade, the most noticeable one being the update
including 3D hydrodynamical simulations to model
the solar photosphere. Up to now, such an analysis has been carried out only
by one group using one radiation-hydrodynamics code.}
{We investigate the photospheric oxygen abundance considering
lines from atomic transitions. We also consider the relationship between
the solar model used and the resulting solar oxygen abundance,
to understand whether the downward abundance revision is specifically
related to 3D hydrodynamical effects.}
{We perform a new determination of the solar photospheric oxygen abundance
by analysing different high-resolution high signal-to-noise ratio
atlases of the solar flux and disc-centre intensity making
use of the latest generation of CO5BOLD 3D solar model atmospheres.}
{We find $8.73 \le \log(N_{\mathrm O}/N_{\mathrm H}) +12 \le 8.79$.
The lower and upper values  represent extreme assumptions on
the role of collisional excitation and ionisation
by neutral hydrogen for the NLTE level populations of neutral oxygen. 
The error of our analysis is $\pm$\,(0.04$\pm$\,0.03)\,dex,
the last being related to NLTE corrections, the first error to any other effect. 
3D ``granulation effects'' do not play a decisive role in lowering
the oxygen abundance.
}
{Our recommended value, considering our present ignorance of
the role of collisions with hydrogen atoms on the NLTE level
populations of oxygen,
is $\log(N_{\mathrm O}/N_{\mathrm H}) = 8.76 \pm 0.07$. 
The reasons which have led to lower O abundances in 
the past are identified as (1) the lower equivalent widths 
adopted, and (2) the choice of neglecting collisions with 
hydrogen atoms in the statistical equilibrium calculations for oxygen. 
}

\maketitle


\section{Introduction}

Oxygen is the most abundant element after hydrogen and helium,
and is
the main product of Type II supernovae. 
In the studies of chemical abundances in stars, and in interstellar
and intergalactic matter, it is common to use the 
solar or solar system abundances as a reference.
Unlike many 
other elements oxygen is volatile 
and is therefore incompletely condensed 
in meteorites, where it is present in the form of oxides.
As a consequence, the only available reference is
the solar \emph{photospheric} oxygen abundance.

In spite of its high cosmic abundance there are not 
many 
atomic oxygen lines 
which are suitable for spectroscopic abundance determinations.
Oxygen in the solar photosphere is mainly present
in the neutral state. The permitted resonance and
low excitation energy lines lie in the 
vacuum ultraviolet, many at wavelength short-ward of the Lyman break.
High-resolution spectra of the Sun at these
wavelengths are not available. However,
even if they were available, the line blending and
heavy saturation of most of these lines
would make them unsuitable for abundance measurements.
In the optical or
near infrared range there are only a few \ion{O}{i}
lines which can be used,
however, all the permitted lines are of 
high excitation and many of these 
are blended.
There are three forbidden lines which arise
from the ground state or very low lying levels,
which, however, are very weak, and they are all
more or less blended with other atomic or
molecular lines.
This makes the determination of the solar oxygen abundance a rather 
complex matter which requires special care. 

Because of its high cosmic abundance oxygen is 
one of the major contributors to the internal opacity of matter
at solar-like chemical composition.
It also plays a special role, together
with nitrogen and carbon, in the energy
production in the Sun.  
This implies that the solar
oxygen abundance has a direct impact on the 
internal structure and evolutionary behaviour of the Sun.
A solar model with reduced abundance of
CNO elements has lower central
density and opacity \citep{watson}
and thus evolves more slowly than a model
with higher abundances.
In the late evolution of the Sun, the CNO cycle
becomes the main source of energy, 
and a reduced abundance of these elements
together with the lower central densities implies
a higher central burning temperature for the Sun.

The oscillatory properties
of the Sun are strongly dependent on the
metal content. In fact, it has been
suggested to use helioseismic data to derive Z \citep{antia}.

The solar oxygen abundance has
undergone several major downward revisions
over the  years, starting from  
the value of \citet[][]{lambert78}, A(O)=8.92\footnote{A(X) $\equiv
\log(N_{\mathrm X}/N_{\mathrm H}) +12$}.
The influential review of \citet{AG89} still provided a value 
which was near to this measurement (A(O)$=8.93 \pm 0.035$), but
nine years later it underwent a downward update by 0.1\,dex
\citep{gs}, then was subject to another downward revision by almost 
0.1\,dex due to \citet{hhoxy}, an other downward revision by \citet{ALA01},
and finally was lowered again
down to the extremely low value of A(O)=8.66 recommended
by \citet{asplund04}. 
The latter result is supported by \citet{melendez04} and \citet{socas07}.
However, the recent result of 
\citet[][A(O)=8.84]{ayres06}, based on CO infrared lines, 
is in counter-trend.

The very low abundance advocated by \citet{asplund04}, together with the
downward revision of the solar abundances of all the other elements by Asplund
and collaborators \citep[][and references therein]{sunabboasp}, has created
serious problems for solar models to explain the helioseismic data
\citep{bahcallA,bahcall,guzik}.

So far, the only three analyses of the solar oxygen abundance which take into
account the effects of photospheric inhomogeneities (granulation effects) 
are those of \citet{hhoxy},
\citet{ALA01}, and that of \citet{asplund04}. The recommended oxygen abundances
differ by almost 0.1\,dex.
According to our analysis based on 3D \cobold\ solar models, the granulation
effects do not lead to a systematic lowering of the oxygen abundance. 
The 3D corrections we find are small and, for the majority of the lines,
positive.

We here present a new and independent determination of the solar
oxygen abundance, with particular emphasis on a latest generation 3D 
\cobold\footnote{http://www.astro.uu.se/$\sim$bf/cobold/index.html}
solar model, which, we believe, helps in understanding
possible underlying systematic errors in the modelling
of the solar photosphere. 
The line by line analysis is presented with full details
in the next sections. However, the paper is organised
in such a way that the reader not interested in such details
may directly skip to the discussion and conclusions in Section
\ref{disc_concl}.


\section{Atomic data}

The atomic data on oxygen lines is reviewed in detail in section 2.1 of
\citet{asplund04} (hereafter A04). For the lines in common we adopt the same
data and we refer the reader to that paper for further details.  In addition
to the atomic lines used by A04 we make use of two IR lines: the 1316\pun{nm}
line for which we adopt the \loggf~ of \citet{biemont91}, and the 1130\pun{nm}
line for which we use the NIST value of \loggf\,=\,0.076, the same value used 
by \cite{hhoxy}.  The adopted atomic data are summarised in
Table~\ref{atomicdata}.

A comment on the collisional broadening constants~$\log(\gamma_6/N_\mathrm{H})$
given in Table~\ref{atomicdata}: in our 3D spectrum synthesis computations we
payed particular attention to the Van der Waals broadening of the oxygen lines
in order to obtain as realistic line profiles as possible. While this
broadening process is important for the \textit{strength} of strong lines it
also plays a role for the \textit{shape} of weak lines. From that perspective
we deemed appropriate to provide the actual values which we employed. Whenever
available we preferred Van der Waals broadening constants obtained from the
theory of \citet{abo1,abo2,abo3} and summarised in \citet{abo4}.  Hereafter,
we shall collectively refer to the description as ``ABO theory''.  
ABO theory predicts a slightly different temperature dependence of the 
broadening effects than classical theory. It is
customary to give the broadening constants for a Maxwellian (relative) velocity
distribution at $10^4$\pun{K}. We did not implement the ABO theory in our 3D 
spectrum synthesis code, but instead opted to ignore the slight differences 
in the temperature dependences and to choose a reference temperature of 
5500\pun{K} at which we evaluated the broadening constants from ABO theory.
This ensures that our classical implementation provided the ABO value at 
this temperature. The reference temperature was chosen as characteristic for 
the lower photospheric layers of the Sun. This is the reason why we give the 
broadening constants in a non-standard fashion with reference to 5500\pun{K} 
in Table~\ref{atomicdata}.

Alternatively, we have computed Van der Waals broadening constants from
the approximation implemented in the Kurucz routine WIDTH \citep[see][]{ryan}.
In short, the WIDTH approximation is the following: 
$C_6 = (\alpha e^2)/h\,(45-S)/Z$, where
$S$ is the number of electrons in the ion under consideration and
$Z$ is the nuclear charge (for neutral oxygen $S$=8 and $Z$=8). 
Then $\gamma_6/N_H = 17C_6^{2/5}v^{3/5}$,
where $v$ is the average relative velocity of the atom with respect to 
the perturber \citep[][Eq. 11.28]{gray}. The results are listed in 
Table\,\ref{atomicdata}.

\begin{table}
\caption{\label{atomicdata}
Atomic parameters of the oxygen lines included in our analysis. 
The table columns give the
wavelength, the transition, the oscillator strength, the excitation 
potential of the lower level, and two choices of the Van der Waals 
broadening constant considered: ``ABO'' refers to results 
from ABO theory, ``WA'' to the WIDTH approximation
(see text for details).
}
\begin{center}
\begin{tabular}{rrrrrr}
\hline\noalign{\smallskip}
\multicolumn{1}{c}{Wavelength}& 
\multicolumn{1}{c}{Trans.}& 
\multicolumn{1}{c}{\loggf}  & 
\multicolumn{1}{c}{$\chi _{\rm lo}$} & 
\multicolumn{2}{c}{$\log(\gamma_6/N_\mathrm{H})$ @ 5500\pun{K}} \\
\multicolumn{1}{c}{in air [nm]} & & &
\multicolumn{1}{c}{[eV]}  &  
\multicolumn{2}{c}{[s$^{-1}$ cm$^3$]}\\
 & & & &   
\multicolumn{1}{r}{ABO} & 
\multicolumn{1}{r}{WA~}\\
\noalign{\smallskip}\hline\noalign{\smallskip}
 630.0304     & $^3$P$-^1$D  & --9.717 &  0.000 &        & --8.362\\
 636.3776     & $^3$P$-^1$D  &--10.185 &  0.020 &        & --8.364\\
 615.81\hspace*{3mm}
              & $^5$P$-^5$D  & --0.296 & 10.741 & --6.948& --7.246\\
 777.1941     & $^5$S$-^5$P  &   0.369 &  9.146 & --7.569& --7.741\\
 777.4161     & $^5$S$-^5$P  &   0.223 &  9.146 & --7.569& --7.741\\
 777.5390     & $^5$S$-^5$P  &   0.001 &  9.146 & --7.569& --7.741\\
 844.6759     & $^3$S$-^3$P  &   0.014 &  9.521 & --7.511& --7.709\\
 926.66\hspace*{3mm}
              & $^5$P$-^5$D  &   0.825 & 10.741 & --7.334& --7.489\\
1130.2377     & $^5$P$-^5$S  &   0.076 & 10.741 & --7.141& --7.565\\
1316.4\hspace*{4.5mm}
              & $^3$P$-^3$S  & --0.033 & 10.99  & --7.108& --7.555\\
\noalign{\smallskip}\hline\noalign{\smallskip}
\end{tabular}
\end{center}
\end{table}


\section{Codes and atmospheric models}

Our analysis is based on a time-dependent 3D hydrodynamical model of the solar
photosphere, as well as on a number of standard 1D hydrostatic models. The
3D model was computed with the \COBOLD\ code 
\citep{Freytag2002AN....323..213F,Freytag2003CO5BOLD-Manual,Wedemeyer2004A&A...414.1121W}. 
An overview of the numerics, aspects of the methodology of its
application, and  basic validity tests can be found in \citet{solarmodels}. 

The 3D model used has a box size of $5.6\times 5.6\times 2.27\,{\rm
Mm}^3$, a resolution of $140\times 140\times 150$ grid points, and spans a
range in optical depth of about $-6.7<\log\tau_\mathrm{Ross}<5.5$.  The
multi-group opacities used by \COBOLD\ are based on monochromatic opacities
stemming from the MARCS stellar atmosphere package \citep{marcsa,marcsb}
provided as function of gas pressure and temperature with high wavelength
resolution. For the calculation of the opacities solar elemental abundances
according to \citet{gs} are assumed, with the exception of CNO for which values
close to the recommendation of \citet{sunabboasp} are adopted
(specifically, A(C)=8.41, A(N)=7.80, A(O)=8.67).  
The monochromatic opacities are binned into representative opacity groups
\citep{nordlund1982,ludwig1992,ludwig1994,voegler2004}. The solar \COBOLD\
model applied here uses twelve bins; to our knowledge this is the highest
wavelength resolution hitherto applied in a 3D stellar atmosphere model.

The equation of state takes into consideration the ionisation of hydrogen
and helium, as well as the formation of H$_2$ molecules according to
Saha-Boltzmann statistics. It ignores the ionisation of the metals,
which is of minor importance ($<$1\,\%) for the bulk thermodynamic
properties. Relevant thermodynamic quantities -- in
particular gas pressure and temperature -- are tabulated as a function of
gas density and internal energy. To ensure accuracy in higher derivatives
(like the adiabatic sound speed) analytical first derivatives are tabulated 
as well.

Note that the tabulated EOS is used only for the computation 
of the bulk thermodynamic quantities (like gas pressure, temperature,
sound speed, specific heats) needed in the hydrodynamical simulations.
The electron pressure is not relevant in this context.
In contrast, for the \emph{post facto} spectrum synthesis from selected 
snapshots, a precise calculation of the electron pressure is important.
A more detailed LTE EOS, including the formation of H$^-$ and the multiple
ionisation of all individual metals, is employed for this purpose.

We emphasise that our finally obtained oxygen abundance is not used to
alter the 3D model atmosphere consistently. Besides the substantial
computational efforts we believe that the feedback would be very small, in
particular since oxygen is not a significant contributor of electrons. We
tested the feedback in the spectral synthesis: changing the oxygen abundance
by 0.1\,dex or changing the oscillator strength by 0.1\,dex gave line 
strengths of the 630\,nm forbidden oxygen line differing by only 0.1\,\%. 

\begin{figure}
\begin{center}
\resizebox{\hsize}{!}{\includegraphics[draft = \draftflag]
{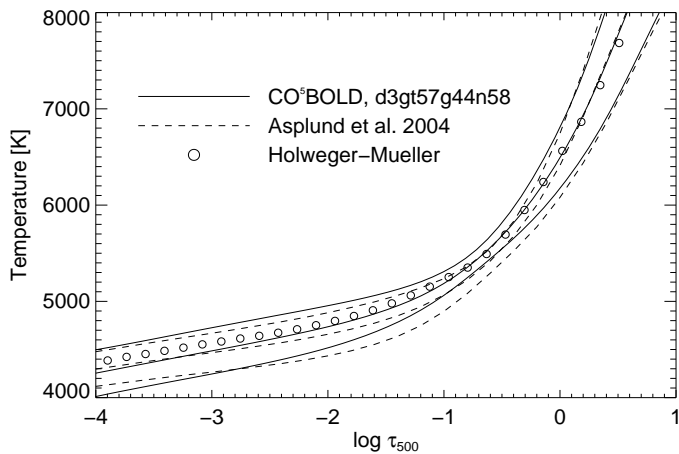}}
\end{center}
\caption[]{Comparison of the temperature structures of the \COBOLD\ model
  used in this work, the 3D model of A04, and the 
  Holweger-M\"uller model. For the 3D models the RMS fluctuations 
  (thin lines) around the mean (thick lines) are indicated.
}
\label{f:coboldasplund}
\end{figure}

For statistically representing the solar atmosphere, we selected 19 snapshots
equidistantly spaced in time covering in total 1.2\,h of solar time. The
snapshots are separated sufficiently in time so that their flow patterns are
largely uncorrelated. They have been further selected to have statistical
properties close to the properties of the whole ensemble of available snapshots. We restricted
ourselves to a sub-set of the available 3D data to keep the computational
burden manageable.
Here, in Fig.\,\ref{f:coboldasplund}, we only show a comparison of the average \COBOLD\ 
model with the 3D model used by A04 in their re-determination of the solar oxygen
abundance. The \COBOLD\ model was averaged on surfaces of equal Rosseland
optical depth. The Rosseland scale was subsequently transformed to an optical
depth scale at 500\,nm in the continuum to allow a direct comparison to the model
published by A04. This is not exactly identical to an averaging on
$\tau_\mathrm{500\,nm}$ surfaces as done by A04 but we consider the involved
inconsistencies immaterial for the present discussion. For further comparison
and benchmark, we added the empirical Holweger-M\"uller  solar
model (\citealt{hhsunmod, hmsunmod}, hereafter HM) to the plot. In the low photospheric layers
($-0.7<\log\tau<0$), the average \COBOLD\ stratification follows the HM model
rather closely. As a consequence, the \COBOLD\ model reproduces the observed
solar centre-to-limb variation in the continuum remarkably well (see \citealt{solarmodels}). 
In this region the A04 model has a steeper
temperature gradient and becomes gradually cooler than the HM model.  Towards
lower optical depth, the \COBOLD\ model becomes increasingly cooler than the
HM model, while the A04 model approaches the HM stratification due to the
different mean temperature gradients of the models.  In any case, both
hydrodynamical models stay systematically cooler than the HM model.  In the
higher photosphere ($\log\tau < -2$) the \COBOLD\ model exhibits larger
temperature fluctuations than the A04 model.

It is unclear which of the 3D models provides the better representation of the
solar photosphere. Nonetheless, we think the differences seen in
Fig.\,\ref{f:coboldasplund} provide an estimate of the systematic uncertainties
which are present in the current generation of hydrodynamical model atmospheres.

As reference we used also several 1D LTE plane parallel
solar model atmospheres:

\begin{enumerate}

\item
An ATLAS model computed
by Fiorella Castelli\footnote{http://wwwuser.oats.inaf.it/castelli/sun/ap00t5777g44377k1asp.dat} 
with version 9 of the ATLAS code and the solar abundances of \citet{sunabboasp}.
With respect to other ATLAS solar models this has been computed with
an Opacity Distribution Function explicitly computed for
the solar chemical composition of \citet{sunabboasp}, 
a micro-turbulent velocity of 1\,\kms, and a mixing-length parameter $\alpha_{\rm MLT}$=1.25.

\item
A solar MARCS model\footnote{http://marcs.astro.uu.se/}, also
computed for a micro-turbulent velocity of 1\,\kms, but with 
$\alpha_{\rm MLT}$=1.5 \citep{marcsb,marcsa}.

\item
The Holweger-M\"uller semi-empirical solar model \citep{hhsunmod, hmsunmod}.

\item
An \emph{average} 3D model denoted by \mD, generated by 
\linfor\footnote{http://www.aip.de/\~mst/Linfor3D/linfor\_3D\_manual.pdf};
this provides a  1D atmospheric structure
by averaging each 3D snapshot over surfaces of equal
(Rosseland) optical depth (cf.\ Fig.\,\ref{f:coboldasplund}). 

\item
A 1D hydrostatic model which employs the same micro-physics as CO5BOLD,
but uses the mixing-length theory to treat convection
(hereafter \xx, see  \citealt{zolfito} for details).
We fixed the mixing-length parameter to $\alpha_{\rm MLT}$=1.0 
and considered two micro-turbulence velocities,
1.0 and 1.5$\mathrm{km\, s^{-1}}.$
We use this model as a reference for differential 1D - 3D comparison.
\end{enumerate}

The spectrum synthesis on the 3D models is done with 
\linfor. A detailed LTE equation of state is employed to provide
the electron pressure and the number density of all individual atoms and ions.
A limitation of the current version of \linfor\ is that
it   cannot handle
a large number of lines and, furthermore, even if it could,
the computational effort would be prohibitive.
We used Linfor3D to compute
3D as well as 1D synthetic spectra, for direct comparison to the hydro-model.
For the 1D synthesis we also used SYNTHE
\citep{1993KurCD..18.....K,2005MSAIS...8...14K},
in its Linux version \citep{2004MSAIS...5...93S,2005MSAIS...8...61S}
that has no limitation in the number of lines
and therefore allows to include a large number 
($\sim 1000$ for a range of 1\pun{nm})
of atomic and molecular lines in order to take into 
account the blending of very weak lines in great detail.
In addition, we employed WIDTH
\citep{1993KurCD..13.....K,2005MSAIS...8...14K,2005MSAIS...8...44C,
2005MSAIS...8...61S}, to derive 1D abundances directly from measured
 equivalent widths.

The global effect produced by many weak lines is similar
to that of an additional continuous opacity. 
In principle it is possible to do an appropriate
``re-normalisation'' of the spectrum and compare this
to a synthetic spectrum computed without inclusion of these
very weak lines.
We tested this procedure 
with 1D synthetic spectra and found it to be
acceptable, at least as a first approximation, introducing
errors in the derived abundance of less than 0.01 dex.
Therefore our approach has been
to use this technique for the 3D analysis, keeping
in the computation only a minimal number of lines ($n < 10$).

Departures from LTE are non-negligible for several permitted \ion{O}{i}
lines, most noticeably for the 777\pun{nm} triplet.
At present our 3D-NLTE line formation code
by Steffen \& Cayrel \citep[see][]{cayrel07} is not
capable of treating oxygen, and we defer full 3D-NLTE
computations to a future paper.

\begin{figure}
\resizebox{\hsize}{!}{\includegraphics[draft = \draftflag]
{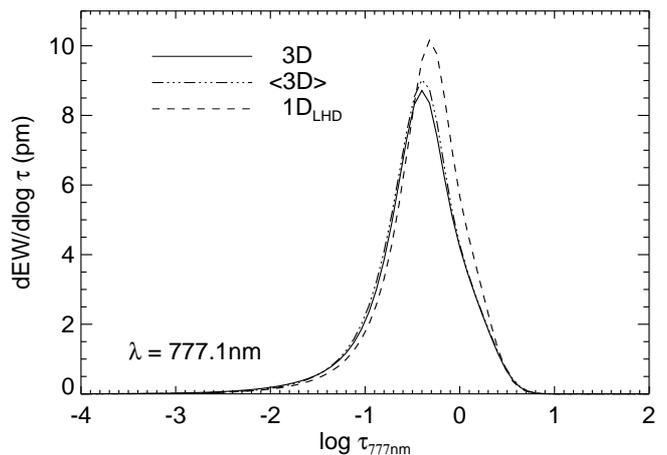}}
\caption[]{The equivalent width contribution function, 
\mbox{dEW/d$\log \tau$},
for the 777.1\pun{nm} line at disc-centre, calculated from the
3D model (solid line), the \mD\ model (triple dot-dashed), and the \xx\ model
(dashed). Following \citet{magain86}, the contribution function is
defined such that integration over $\log\tau_\lambda$ gives
the line equivalent width.}
\label{confoi7771}
\end{figure}

However, in the case of \ion{O}{i} lines 1D computations
should yield a reliable estimate of departures from LTE.
In fact the line formation range in the solar photosphere 
is very similar in 3D and \xx\ models, respectively, 
for the oxygen triplet (see Fig.\,\ref{confoi7771}), and also
for the other infrared oxygen lines.
This is not necessarily valid for other elements and other stellar 
models. For instance, 3D-NLTE and 1D-NLTE contribution functions
of the 670.8\pun{nm} $^7$Li line are very different for a metal-poor 
model at \Teff=6240K, \logg=4.0, \moh=--2.0 \citep{cayrel07}.
Even if the \emph{mean} vertical stratification of oxygen is very similar
in 1D and 3D, horizontal inhomogeneities could introduce a difference between
1D and 3D NLTE corrections. The error due to the 1D approximation needs
to be quantified by future detailed 3D-NLTE calculations. 

The similarity between NLTE corrections computed in 1D and 3D is
confirmed by the results in Table~2 of A04
We suspect
that the two corrections would have been even closer if the authors
had used as a background model for their 1D-NLTE computations
a model with the mean temperature structure of their 3D model.

For the present paper we decided to compute 1D-NLTE corrections
based on the \mD\ model (rather than on \xx), and to
use the Kiel code \citep{SH} together with the oxygen
model atom described in \citet{paunzen}.
We also compare our NLTE corrections with those
computed by \citet{hhoxy} and A04.


\section{Observational data}

There are several high resolution, high signal-to-noise
ratio (S/N) spectra both of the solar flux and of 
the disc-centre intensity, and we decided to make use
of all of them. The choice of using more than one solar atlas
is due to the fact that, even if S/N is very high for solar spectra,  
the abundances obtained from different spectra are not always in 
agreement within one $\sigma$, highlighting the existence of systematic 
differences which are not fully understood.

\begin{enumerate}

\item
The data we will refer to as ``Kurucz Flux'' spectrum
is based on fifty solar FTS
scans taken by James Brault and Larry Testerman at Kitt Peak between
1981 and 1984, with a spectral resolving power $\lambda/\Delta \lambda$
of the order of 300\,000 and a S/N of around 3\,000, both varying
from range to range. Further details can be found in  \citet{kuruczflux}.

\item
The ``Neckel Flux'' and ``Neckel Intensity'' spectra refer to
the \citet{neckelobs} absolutely calibrated FTS spectra obtained at 
Kitt Peak in the 1980ies, covering the range 330\pun{nm} to 1250\pun{nm} for 
full-disc and disc-centre, respectively \citep[see also][]{neckel1999}.  
The  spectral purity ($\Delta\lambda$) ranges from 0.4\pun{pm} at 
330\pun{nm} to 2\pun{pm} at 1250\pun{nm} (825\,000 $>$ 
$\lambda/\Delta\lambda$ $>$ 625\,000) .

\item
\citet{delbouille} provide a disc-centre intensity spectrum in the 
range 300\pun{nm} to 1000\pun{nm} observed from the 
Jungfraujoch\footnote{http://bass2000.obspm.fr/solar\_spect.php}.
At the same web site also an infrared disc-centre 
intensity spectrum, observed from Kitt Peak by \citet{delbouilleir},
is available; this spectrum covers the range
from 1000\pun{nm} to 5400\pun{nm}. 
In the following we shall refer to both atlases, for
brevity, as ``Delbouille Intensity''.

\item
\citet{wallace} present disc-centre intensity FTS spectra
obtained at  Kitt Peak by \citet{pierce}.
We refer to this data as ``Wallace Intensity''.

\item
For the infrared lines at 1130\pun{nm} and 1316\pun{nm} we analysed 
Kitt Peak FTS spectra from the Digital Archive of the National 
Solar Observatory as reduced by one of the authors (TRA).

\end{enumerate}

The light emerging from the Sun is subject to a gravitational
redshift of 636 ms$^{-1}$; of the atlases we employ, only 
the Kurucz Flux has been corrected for this effect.
In addition, one should take into account the effect of the Earth's
rotation at the time the spectrum was acquired. Again 
only the Kurucz Flux has been carefully corrected for this 
effect.  In principle 
the  wavelength precision of the FTS scans is extremely high;
that is, wavelengths measured at one end of the scan
are very close relative to wavelengths measured at the other end of the scan.
In the terrestrial reference frame the scans have high accuracy, 
since they are slaved to a laser.
However, the absolute accuracy depends on how well one has corrected
for all the other effects.
Then there is the additional issue
of how well the laboratory wavelengths of the oxygen lines
themselves are known.
Considering in addition that the convective line shifts predicted
by our hydrodynamical simulations are not perfectly accurate
(we estimate the accuracy to be of the order of 50 ms$^{-1}$), 
we decided to introduce the wavelength shift between observed
and synthetic spectrum as free parameter in the spectral line 
fitting procedure described below.

Considering the six oxygen lines at 630, 636, 777 and 844\pun{nm},
we measure the following wavelength shifts, with respect to
the laboratory wavelengths:
for the Kurucz Flux $21\pm 264$\,\ms, 
for the Neckel Flux $388\pm 258$\,\ms and
for the Delbouille Intensity $344\pm 273$\,\ms.
Our results are in good agreement with \cite{allendeshift}
data who present a $\sigma$ of 238\,\ms with an average of 286\,\ms
for iron lines in the Neckel Flux spectrum.

It is beyond the purpose of the present paper
to investigate the convective shifts of oxygen lines. 
What we wanted to point out are the systematic differences
in absolute wavelengths
between the various solar atlases. This makes it mandatory
to leave the wavelength shift as a free parameter when performing 
line profile fitting.


\section{Data analysis}

To derive oxygen abundances, we measured the equivalent width (EW) of
the oxygen lines and, when possible, fitted the line profile.  EWs
were computed by direct numerical (trapezoidal) integration 
or by integration of a fitted Gaussian or Voigt profile, using
the IRAF task {\tt splot}. The results are summarised in
Table~\ref{oabun}.

All line profile fitting has been performed through $\chi ^2$ minimisation
using a code described in  \citet{eczolfo}.
Besides the oxygen abundance, also the level
of the continuum and a wavelength shift are left as free parameters
in the fitting procedure.
For 1D models, the macro turbulence has to be fixed {\em a priori}.
The whole line profile or only a selected range can be fitted,
as was done, for example, for the 926\pun{nm} line because of the 
presence of a residual of a telluric line in one of the wings.

A change in the Van der Waals broadening constant $\gamma_6$,
input to the line formation code, can produce lines with different
shapes (more or less broadened) but also of different EW.
For instance if we change the $\log\gamma_6$ of the 636\pun{nm} [OI] line
from the standard input value for \linfor\ $-8.145$
to the value of $-8.364$
provided by the WIDTH approximation \citep[][]{ryan}, 
the difference in EW is 0.2\%.
Considering the oxygen triplet, the EW increases by
4\% to 5\% (depending on the oxygen abundance) 
if we change the $\log\gamma_6$ from $-7.57$ as derived
from ABO theory 
to  $-7.74$ as obtained from the WIDTH approximation.
We adopted the following strategy: for the transitions for
which the broadening data were available from ABO 
we computed the
$\log\gamma_6$ according to the ABO theory; when not available,
we used the WIDTH approximation which is the one used also inside
the SYNTHE code. We point out that, contrary to naive expectations, 
the broadening effect is not completely negligible for
weak lines, as also pointed out by \citet{ryan}. 
In the 615\pun{nm} line, for instance, a change
in $\log\gamma_6$ of 0.16 dex, from $-7.164$ to $-7.004$, 
produces, for an oxygen abundance of 8.66, a difference in EW of 1\%.
In contrast, the [OI] lines are insensitive to the value
adopted for $\log\gamma_6$.

Concerning the quadratic Stark broadening, even strong lines
such as the oxygen triplet lines are not so sensitive to this parameter.
In fact a change by a factor 2 in the $\gamma_4$ produces
in the strongest line of the triplet a change
of $^{-0.01}_{+0.006}$ dex in the oxygen abundance 
according to WIDTH combined with the ATLAS model.
Even the shape of the line is very little affected by this change,
only the wings being appreciably different in the synthetic spectra.

Since the emphasis of this work is on the 3D analysis, we analysed 
with \cobold+\linfor~ all the  observed spectra.
For comparison we considered the 1D computations for EWs for
all the  observed spectra,
but we mostly limited the 1D line-profile fitting to the
Kurucz Flux, when available, or to the Delbouille Intensity.

As in \citet{zolfito}, we define the ``3D corrections'' as 
the difference A(O)$_{\rm 3D}$ -- A(O)$_{\rm 1D_{\rm LHD}}$, 
because the 3D effect is best captured when comparing
the 3D and the  \xx\ models (which share the 
same input physics) and using the same line formation code,
\linfor. Note that the resulting ``3D corrections'' depend
on the choice of both the mixing-length parameter, $\alpha_{\rm MLT}$,
and the micro-turbulence parameter, $\xi$, necessary for defining 
the 1D model. The micro-turbulence is important only for the 
stronger lines, in our case the triplet lines.
In this study we considered two values of the micro-turbulence,
$\xi$\,=\,1.0 and 1.5~\kms (see Table~\ref{oabun}). 
Apart from the triplet lines, the difference
in abundance when considering the two $\xi$ is at maximum 0.02\,dex.
For the triplet a different choice in $\xi$ is translated to
a change in the 1D abundance by up to 0.04\,dex.
Higher micro-turbulence produces stronger lines in synthetic spectra,
so that the oxygen abundance obtained is lower.
In the present work we gave preference to $\xi$=1.0~\kms.
Concerning the \xx\ model for the mixing-length parameter we took
$\alpha_{\rm MLT} = 1$. Most of the lines are formed in layers which are not
strongly affected by convection, and hence the choice of the mixing-length parameter.
However, this is not true for the 615\pun{nm} line, which is formed in deep
photospheric layers.
For testing this aspect, we considered three representative lines: the 
strongest line of the triplet at
777.1\pun{nm}, the stronger forbidden line at 630\pun{nm}, and the line at 615\pun{nm}.
We changed the mixing-length parameter by $^{+1.0}_{-0.5}$ with respect to our
preferred
value of 1.0. With this change in $\alpha_{\rm MLT}$ the change in the 
\xx\ oxygen abundance (for fixed equivalent width in the synthetic flux 
spectra)
for the forbidden and the triplet line is small, $^{-0.003}_{+0.011}$ for the
forbidden line, and $^{-0.003}_{+0.018}$ for the triplet line.
In contrast, for the 615\pun{nm} line we find a change in abundance of
$^{-0.045}_{+0.080}$. This clearly indicates that for some lines the 3D-1D
corrections dependent on the mixing-length parameter assumed in the 1D
comparison model.

For completeness, we shall also provide differences using other 
1D models, either with \linfor\ or with SYNTHE. We point out, 
however, that these latter differences depend also on code
specific disparities such as different opacities, partition functions,
radiative transfer schemes, etc, and hence serve mainly to illustrate
the magnitude of systematic modelling uncertainties.


\subsection{Corrections for departure from local thermodynamic equilibrium}

Besides the fact that \citet{hhoxy} computed 1D non-LTE
and A04
3D-non-LTE, the differences in the two NLTE computations 
is the way in which
excitation and ionisation by inelastic collisions
with hydrogen atoms is accounted for.
\citet{hhoxy} accounts for them using the formalism
of \citet{SH}, which is a generalisation of that of \citet{Drawin}.
A04 instead totally ignore them. 

The two reasons which favour ignoring collisions
with hydrogen atoms are (i) the larger cross-section
for collisions with electrons and (ii) the fact that
electrons have much higher velocities, thereby making
collisions with $e^-$ far more frequent. 
On the other hand, the ratio of hydrogen atoms
to electrons is often as high as $10^4$ or even $10^5$.
Under such circumstances it appears difficult to deny 
the role of collisions with hydrogen atoms.

In our solar model, $N_\mathrm{H}/N_\mathrm{e}$ is almost flat at $10^4$ in the 
outer layers, it is about $0.8\times10^4$ at
$\log\tau_\mathrm{Ross} = -0.5$ and then drops dramatically
to  $\sim 70$ at $\log\tau_R = 1$.
One should therefore expect that the role of collisions 
with hydrogen atoms is different for lines having
different line formation depths.

Since the \citet{SH} formalism is the only
practical way to include the effects of collisions
with hydrogen atoms,
it has become customary in NLTE codes to adopt it
and to introduce the scaling factor ${\rm S_H}$. 
A zero value for this scaling factor is equivalent
to ignoring the process. 
For some species, like Na and Li, for which laboratory data
exist, it is known that the classical Drawin formula \citep{Drawin}
leads to an overestimate of the collisional rates 
\citep[see][and references therein]{asplund04}.
Unfortunately, no laboratory data exists for collisions of neutral
hydrogen with oxygen. Inspection
of the literature shows that very different choices for
${\rm S_H}$ are made for different species.
For example \citet{gehren} advocate ${\rm S_H}=5$ for \ion{Fe}{i}
in the Sun. 

\citet{carlos04} have used the centre-to-limb variations of
the 777\pun{nm} \ion{O}{i} triplet to constrain the role
of collisions with hydrogen atoms.
Their $\chi^2$ analysis of synthetic NLTE profiles computed
from their 3D hydrodynamical simulation favours
${\rm S_H}=1$. 

In several papers the Kiel group has adopted
${\rm S_H}=1/3$ \citep{solanki,STUR90,STUR91},
mainly based on unpublished work of Steenbock.
In the present paper we present NLTE corrections
for ${\rm S_H}=0,1/3,1$ and discuss the implications.


\subsection{Error estimates}

We here consider the possible errors that affect the analysis
of the oxygen abundance in each line.
We will take into account for each line  the following errors:
``\loggf~error'', ``broadening theory error'',
``blending error'', ``NLTE correction error'', ``statistical error''.

\begin{enumerate}

\item
The  ``\loggf~error'': the uncertainty of \loggf~ translates directly
into an error of the oxygen abundance.
Since \citet{SZ} do not report an error estimate of \loggf\ for
the two forbidden lines, we considered as the error in \loggf~ 
the difference between the value given by \citet{SZ}
and the value used by \citet{lambert78}, which has been widely used prior
to the \citet{SZ} computations. 
For the other oxygen lines we used the error of the $f$-value as listed
in the NIST Atomic Spectra 
Database\footnote{http://physics.nist.gov/PhysRefData/ASD/index.html, 
\citet{Ralchenko}}.

\item
The ``broadening theory  error'': in our analysis we applied,
whenever possible, the ABO broadening theory. However, for some lines
the use of the WIDTH approximation would result in a different abundance.
We therefore consider the difference in the derived abundance using the 
two different broadening theories as an upper limit of the uncertainty 
associated with the broadening theory itself.

\item The ``blending error'': some oxygen lines are blended with other lines,
either stellar or telluric.
Uncertainties in the atomic data of the blending line(s) as well as
uncertainties in the abundances of the involved species introduce additional
errors which propagate to the final oxygen abundance.
In some cases, the presence of a neighbouring line, influencing the wings of
the oxygen line in question, can produce an uncertainty in the placement of
the continuum. This is the case, e.g., for the 844\pun{nm} line. These two
kinds of uncertainties associated with blending were considered together.
The error due to uncertain continuum placement, because of blends in the
line wings, has been estimated through successive trials, identifying the
``highest acceptable'' and ``lowest acceptable'' continuum level.  There is
clearly some degree of subjectivity in this estimate. However, we were
unable to devise a more objective approach.
Concerning the telluric absorption an uncertainty is introduced when
the line is subtracted from the spectrum.

\item
The ``NLTE correction error'':
the choice on the value to assign to the H-atom collision
scaling factor, ${\rm S_H}$, changes the NLTE correction.
As the related uncertainty we use the difference in the 
(logarithmic) NLTE abundance correction between S=0 and S=1/3.

\item
The ``statistical error'':
is simply the standard deviation of the abundances
derived from the different observed spectra. 

\end{enumerate}

The error related to the S/N, estimated from the Cayrel
formula \citep{cayrel88}, is completely negligible
in comparison to the errors listed above.
The differences between different solar atlases
cannot be due to the random noise in the spectra, 
which is very low for all the data considered here, but
must be related to some undetected systematic effects.

The uncertainties are summarised in Table~\ref{finaboxy}.
The column ``total error'' of this table considers the global
uncertainty that we attribute to each line; this value is the 
quadratic sum of all the uncertainties.
The error related to the NLTE computations is not included in
Table~\ref{finaboxy} because it is considered as a systematic error.
Details of the derived NLTE corrections are instead listed in 
Table~\ref{nltecorr}.


\subsection{\label{s:linedetails} Detailed analysis of individual transitions}

In this Section we give details of the analysis of each of the considered
transitions.  
The results are summarised in Tables~\ref{oabun} and \ref{finaboxy}.
Initially, we considered also the 557\pun{nm} forbidden line,
but discarded it because both the Delbouille Intensity and the Kurucz Flux
show an asymmetry in the line profile, due to the presence of C$_2$ lines, as
pointed out by \citet{lambert78}. Unfortunately the molecular data for these
C$_2$ lines is still not of sufficient quality to allow a reliable synthesis
of the profile.

\subsubsection{[OI] 630 nm}

\begin{figure}
\resizebox{\hsize}{!}{\includegraphics[clip=true,angle=0]{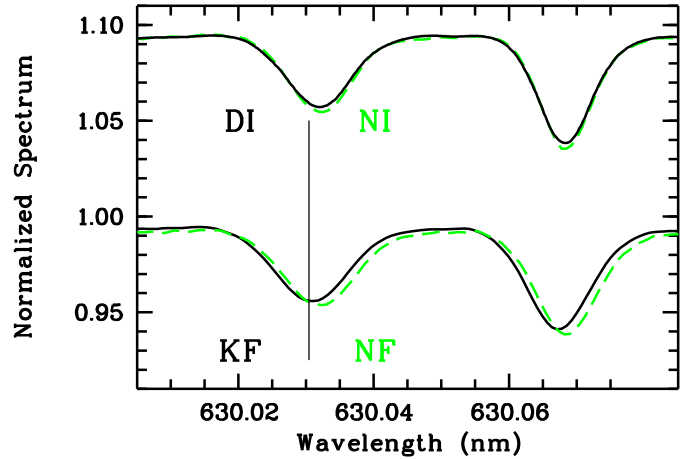}}
\caption{The four observed spectra we used for the analysis of the
[\ion{O}{i}] 630\pun{nm} line.
The Kurucz Flux spectrum, KF, and the Neckel Flux, NF, and Neckel Intensity, NI, spectra
here plotted are the normalised spectra provided by \citet{kuruczflux} and
\citet{neckel1999} respectively; 
the Neckel Intensity, NI, is shifted up by 0.1
for clarity, while the Delbouille Intensity, DI, has been normalised
and then shifted up by 0.1. The vertical line indicates the laboratory 
wavelength of the oxygen line.
We recall that KF is corrected for gravitational redshift while NF is not,
explaining the large wavelength differences between the two flux atlases}.
\label{oi630}
\end{figure}

\begin{figure}
\resizebox{\hsize}{!}{\includegraphics[clip=true,angle=0]{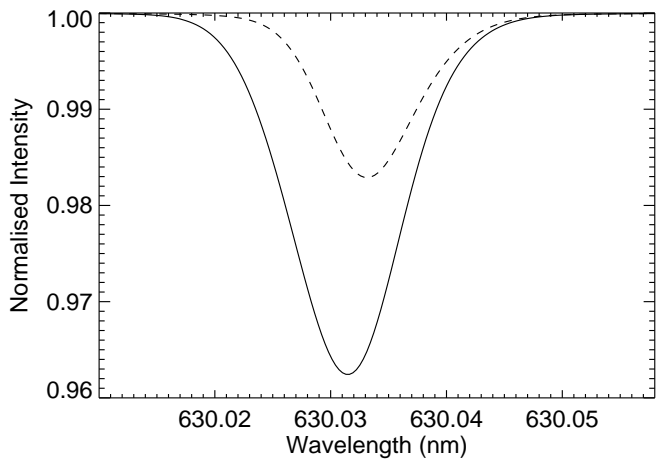}}
\caption{The contribution of \ion{Ni}{i} line (dashed line) with A(Ni)=6.25
is compared to the combined [\ion{O}{i}]+\ion{Ni}{i} blend (solid line)
for disc-centre according to the \linfor\ computation with A(O)=8.66.
}
\label{o630blend}
\end{figure}

This weak line (see Fig.\,\ref{oi630}) arises from the ground level
and is immune to departures from LTE, so it has 
always been considered a primary indicator of the photospheric
oxygen abundance. However, it is blended with a \ion{Ni}{i} line and
lies next to two stronger lines\footnote{\ion{Si}{i} line: 
$\lambda$ = 629.9599\pun{nm},
\loggf = $-1.50$, $\chi_{\rm lo}$=5.948\pun{eV}; 
\ion{Fe}{i} line:
$\lambda$ = 630.1501\pun{nm}, \loggf = $-0.72$,
$\chi_{\rm lo}$=3.654\pun{eV}},
whose wings can influence the EW of the oxygen line if not
properly taken into account.
In Fig.\,\ref{o630blend} the contribution of nickel is
compared to the blend [\ion{O}{i}]+\ion{Ni}{i}
according to a 3D synthetic spectrum.

The blending nickel line exhibits a complex isotopic structure
(see Table~\ref{isoni}).
The two main components are $^{58}$\ion{Ni}{i} and $^{60}$\ion{Ni}{i}, 
with an isotopic ratio of 0.38
and $\lambda $ of 630.0335\pun{nm} and 630.0355\pun{nm}, 
respectively (see \citealt{bensby}).
Each component of the nickel blend is assumed to have \loggf=$-2.11$
\citep{johansson}.
The other two even isotopic components, $^{62}$\ion{Ni}{i}
and $^{64}$\ion{Ni}{i}, can be inserted in the line list as well,
even if this does not change significantly the line shape,
assuming that their shift with respect to the other even components
has the same value as the shift between $^{58}$\ion{Ni}{i}
and $^{60}$\ion{Ni}{i} \citep[about 0.002~\pun{nm}, see][]{rosberg}.
\citet{rosberg} studied the isotope splitting of the even components of 
\ion{Ni}{ii} and showed that the splitting is equidistant for 
any of the even components of \ion{Ni}{ii} they studied. 
We assume the same behaviour for \ion{Ni}{i}.

\begin{table}
\caption{\ion{Ni}{i} isotopic splitting.}
\label{isoni}
\begin{center}
\begin{tabular}{rrrrr}
\noalign{\smallskip}\hline\noalign{\smallskip}
\ion{Ni}{i} isotope & \%      & Wavelength & \loggf & \% used\\
\noalign{\smallskip}\hline\noalign{\smallskip}
$^{58}$\ion{Ni}{i} & 68.27 & 630.0335 & --2.11 & 72.34 \\
$^{60}$\ion{Ni}{i} & 26.10 & 630.0355 & --2.11 & 27.66 \\
\noalign{\smallskip}\hline\noalign{\smallskip}
$^{58}$\ion{Ni}{i} & 68.27 & 630.0335 & --2.11 & 69.05 \\
$^{60}$\ion{Ni}{i} & 26.10 & 630.0355 & --2.11 & 26.40 \\
$^{62}$\ion{Ni}{i} &  3.59 & 630.0375 & --2.11 &  3.63 \\
$^{64}$\ion{Ni}{i} &  0.91 & 630.0395 & --2.11 &  0.92 \\
\noalign{\smallskip}\hline\noalign{\smallskip}
\end{tabular}
\\
\end{center}
\end{table}

We adopted A(Ni)=6.25 \citep{gs} and consider it as a fixed parameter 
in the oxygen abundance determination.

The observed solar spectra were renormalised 
in order to remove the silicon and iron contributions and
the EW of the observed spectra were computed with the program
{\tt splot}. The oxygen abundances obtained from the curve-of-growth 
computed with \linfor\ are reported in Table~\ref{oabun}.
Note that the abundance obtained from HM+\linfor\ is only 
\emph{slightly higher} by $\approx$ 0.015 dex. On the other hand,
both 1D ATLAS and MARCS models, in combination with
the WIDTH line formation code, imply  a 
\emph{lower} oxygen abundance by 0.03\,dex.
In this respect both MARCS and ATLAS models 
behave in a way which is similar to the LHD model.
As can be deduced from Table~\ref{oabun},
the difference 3D-\mD ~ is considerably smaller than 
the difference 3D-\xx, meaning that the latter difference
is dominated by the different mean temperature structures
rather than by horizontal temperature fluctuations.

Note that we disagree on the sign of the 3D-1D correction for this
line with what was previously found by A04
whose results
imply a 3D correction A(O)$_{\rm 3D}$ -- A(O)$_{\rm 1D MARCS}$ of
$-0.04$\,dex for the flux, while we find a value of $+0.03$\,dex (see
above).

Our 3D correction for this line is about +0.05\,dex for the flux and +0.07\,dex
for the intensity spectra. The influence of the micro-turbulence is completely
negligible for this line.

From this line, fixing A(Ni)=6.25, the 3D oxygen abundance is 
$8.679\pm 0.014$ if we consider all four spectra. If we consider only
the two flux spectra A(O)$_{\rm 3D}=8.681\pm 0.019$,
and A(O)$_{\rm 3D}=8.676\pm 0.014$ from the two disc-centre spectra.

\begin{figure}
\resizebox{\hsize}{!}{\includegraphics[clip=true,angle=0]{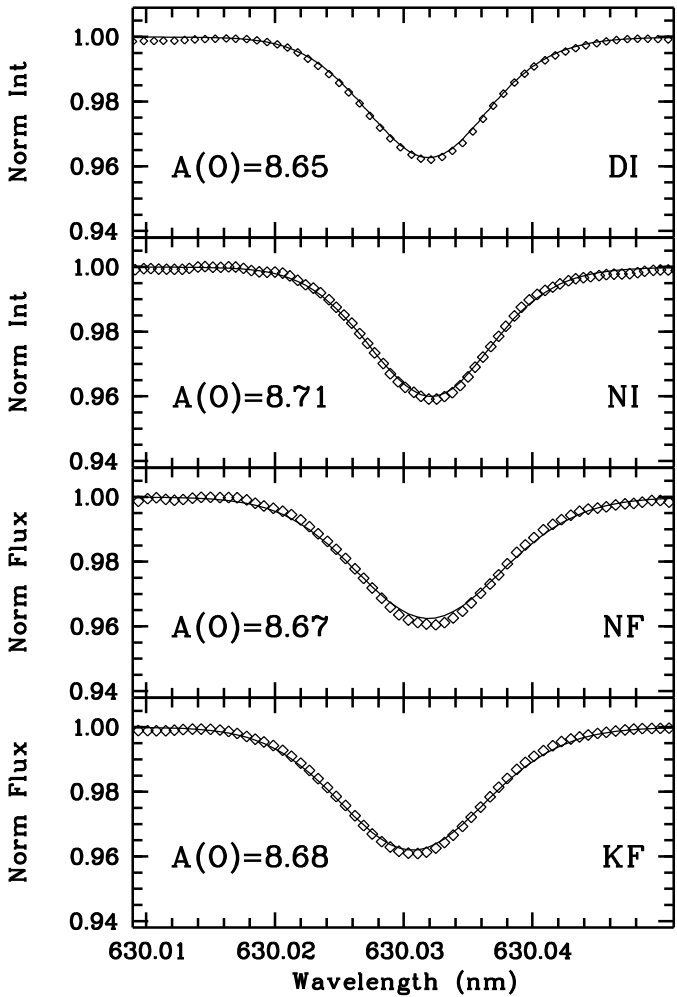}}
\caption{The 3D fit of the [OI] 630\pun{nm} line.
The observed spectra (diamonds, their size indicating the observational
noise level) are over-imposed on the best fit 3D synthetic line profile 
(solid). From top to bottom, the spectra are: Delbouille Intensity (DI), 
Neckel Intensity (NI), Neckel Flux (NF), and Kurucz Flux (KF).
}
\label{fit3d_o630}
\end{figure}

\emph{Fitting} the line profile of the solar intensity and flux
spectra with a \cobold+\linfor\ grid of synthetic spectra, we obtain an
oxygen abundance which is in perfect agreement with the results
obtained from the EWs (Table~\ref{oabun}) within 0.01\,dex 
(see Fig.\,\ref{fit3d_o630}).

The 1D fits, even if not able to reproduce satisfactorily the
line shape, are in agreement with the 3D results, giving further
support to the notion that the 3D correction is small.
The line fit using SYNTHE slightly {\em underestimates} the
oxygen abundance (HM model A(O)=8.67, MARCS model A(O)=8.65
ATLAS model A(O)=8.61), in agreement with the above result based
on matching the EW with 1D models.

Since this oxygen line is blended with the neutral nickel line,
a possible approach is to use a  
multi parameter fit, with oxygen and nickel abundance,
wavelength shift, and continuum level as free parameters.
We attempted this, but the result was not satisfactory because 
wavelength shift and nickel abundance turned out to be
closely correlated parameters, thus producing a range of
degenerate solutions for the best fit. However, if the wavelength
scale can be established by other methods, the degeneracy is
removed and both the nickel and the oxygen abundance can be found
from line fitting \citep[see][and Sect.\,\ref{disc_concl} below]{ayres08}.

Since an error in the parameters of the nickel line
introduces an error of the oxygen abundance,
we computed two additional 3D grids 
where we varied the \ion{Ni}{i} abundance by $\pm 0.1$\,dex
with respect to its fiducial value of 6.25.
Noting that the error of the solar nickel abundance is 0.04\,dex \citep{gs},
and the error of the \loggf~value of this line is
0.06\,dex \citep{johansson}, a variation of  $\pm 0.1$\,dex in the
strength of the nickel line takes into account the possible uncertainties.
Our evaluation of the systematics induced in the oxygen abundance shows
that a difference of $\pm 0.1$ in the nickel abundance produces a 
difference in the oxygen abundance of $^{-0.03}_{+0.05}$ 
(resulting in A(O)=8.66, 8.74 respectively) in the fitting of the Kurucz Flux.
Similar values are obtained for the Delbouille Intensity 
(a $\pm 0.1$ change in nickel implies $\mp 0.06$ change in oxygen abundance).

Our  adopted oxygen abundance for this line comes from the average value
of the four observed spectra, and it is
A(O)=$8.68\pm (0.014)_{\rm stat}\pm (0.15)_{\rm syst}$
(see Table~\ref{finaboxy} for details). 
The systematic error is the quadratic sum of the
\loggf\ error and the error related to blends and uncertainties in 
the continuum placement.
The error due to the broadening theory is completely negligible for this
line, being of the order of $10^{-4}$\,dex.
The final result is \boldmath {\bf A(O)=$8.68\pm 0.15$}. \unboldmath


\subsubsection{[OI] 636 nm}

\begin{figure}
\resizebox{\hsize}{!}{\includegraphics[clip=true,angle=0]{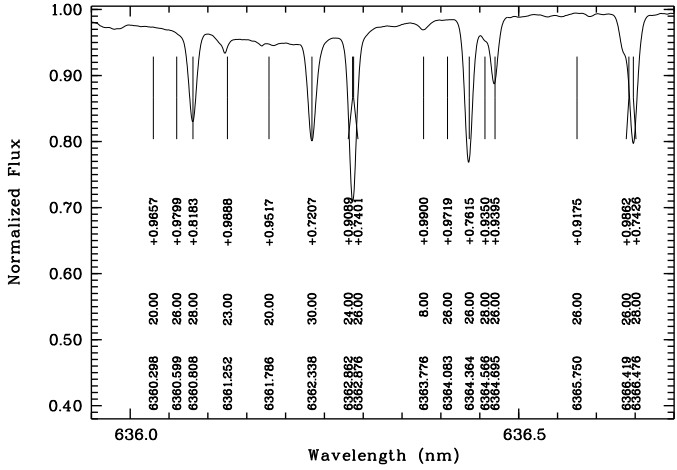}}
\caption{The Kurucz Flux spectrum around the  [OI] 636.378\pun{nm} line
with the identification of the stronger lines.
}
\label{O636line}
\end{figure}

This line lies on the red wing of a \ion{Ca}{i} auto-ionisation line
(see Fig.\,\ref{O636line}). Several CN lines are present in the range. 
For C and N abundances we adopted 8.39 and 7.80 respectively.
Their contribution to the EW of the [OI] 636 line, according to 
our ATLAS+SYNTHE synthesis, is of 0.05\pun{pm} in flux and
0.03\pun{pm} in intensity.
With a \cobold+\linfor\ synthetic spectrum, the CN molecules
happen to come out stronger by 20\% with respect to the 1D
ATLAS+SYNTHE result. 
This  implies a difference of only 0.002\,dex in the
derived oxygen abundance, which is far from being able 
to explain the 0.1\,dex difference between the oxygen abundance
provided by this line (see below) and that of the 
630 nm forbidden line (see Table~\ref{oabun}).

At first we tried to remove the contribution of
all the blending lines by dividing the observed spectrum by a 1D synthetic
spectrum computed with all relevant atomic and molecular
lines (including the \ion{Ca}{i} auto-ionisation line) {\em except} oxygen. 
We obtained an oxygen abundance (A(O)=8.78), inconsistent with what was 
derived from the other forbidden line (A(O)=8.68).
This is at variance with what found by A04,
according to their
analysis, the  difference in abundance is of only 0.02\,dex.
This can stem from the differences in the EWs.
But a difference of 0.06\pun{pm} in EW for this line can hardly
explain the difference in A(O). A differencemust rely on the spectral-synthesis
code used in the two analysis.
Next, we normalised the spectrum following the
profile of the \ion{Ca}{i} auto-ionisation line.
From the EW measured from this normalised spectrum 
we subtracted the contribution due to
CN molecules and obtained the oxygen abundances reported
in Table~\ref{oabun}. Our oxygen abundance for this line relies 
therefore only on the EW approach. 

We point out that the three spectra observed from Kitt Peak (Neckel and
Kurucz) imply a higher oxygen abundance than the one observed from the
Jungfraujoch (Delbouille). 
Moreover, also  for the [OI] 630\pun{nm} line there is a difference
in between the two intensities: in the Neckel Intensity spectrum the line
is stronger than in the Delbouille spectrum, even if the effect is less 
evident.
We considered the hypothesis that the difference 
in height above sea level of the two sites
could have induced a higher telluric contribution in the spectra
taken at the site of lower elevation. But since there is no evidence of 
a telluric contribution from the inspection of spectra taken at different
air mass, we discarded this hypothesis. More likely, the differences
in the spectra might be related to different levels of solar activity at 
the time of the respective observations \citep[cf.][]{livingston2007}.

The fact that the oxygen abundance derived from the two forbidden lines 
is clearly not in agreement is disturbing. Obviously, deficiencies in
the solar model atmospheres cannot be responsible. In fact, 3D and HM
show a similar discrepancy (see Table~\ref{oabun}). Presumably, the problem
is related to our imperfect knowledge of the blend line contributions.
A possible solution is offered by the work of \citet{ayres08}.

Our adopted oxygen abundance for this line, taking into account
all the measurements, is 
$8.78\pm (0.05)_{\rm stat}\pm (0.12)_{\rm syst}$,
which is \boldmath {\bf A(O)=$8.78\pm 0.12$} \unboldmath
(see Table~\ref{finaboxy} for details).


\subsubsection{\ion{O}{i} 615 nm}

\begin{figure}
\resizebox{\hsize}{!}{\includegraphics[clip=true,angle=0]{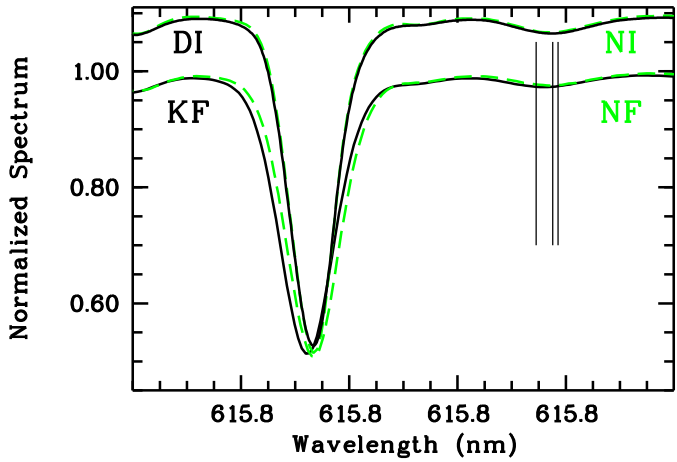}}
\caption{The \ion{O}{i} 615\pun{nm} Mult.~10 lines  in the four observed 
spectra used in this work. Symbols are as in Fig.\,\ref{oi630}. The three 
vertical lines indicate the laboratory wavelength of the lines.
}
\label{oi615}
\end{figure}

These lines of Mult.~10 lie in a very complex spectral region (see Fig.\,\ref{oi615}),
close to a rather strong iron line and in the range
of many CN lines which make the continuum placement
problematic.
The EW was measured using the de-blending capability
of {\tt splot}, with which we accounted for
the presence of the neighbouring \ion{Fe}{i} line.
The EW is very sensitive to the placement of the continuum.
The oxygen abundance is then obtained from matching this EW
with the 3D curve-of-growth.

Our adopted LTE oxygen abundance for this line is
8.63 for the disc-centre spectra and 8.64 for the flux spectra.
For both disc-centre and flux we estimate a global systematic error of 
0.21 dex, while the statistical error is 0.01 for flux and 0.03 for intensity
(see Table~\ref{finaboxy} for details). 
We neglect any possible error related to the molecular
data because we think they are included in the
error associated with the continuum level placement.

The \xx\ results overestimate the oxygen abundance by  
about +0.03\,dex (flux), in other words the 3D correction 
is negative, $-0.03$\,dex. 
Only for this line we find a negative 3D correction, and this is
a consequence of the fact that only in this case
the effect of horizontal temperature fluctuations, 3D-\mD,
is appreciable. This, in turn, is a consequence of the high
excitation potential of this line which implies that it forms in the
deep photosphere and is highly sensitive to temperature fluctuations.

Using the Holweger-M\"uller model, the oxygen abundance is 
even 0.14\,dex higher than the 3D result.
For comparison, the corresponding abundance difference is
0.12\,dex for ATLAS+WIDTH, and 0.06\,dex for MARCS+WIDTH.

\begin{figure}
\resizebox{\hsize}{!}{\includegraphics[clip=true,angle=0]{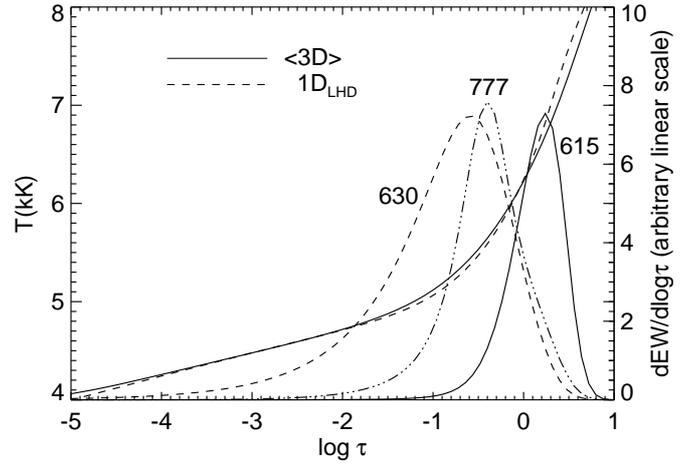}}
\caption{The average temperature profile (left ordinate) of the 3D model
(solid line) and the temperature profile of the \xx\ model (dashed) are
depicted as a function of Rosseland
optical depth. The EW contribution function of the disc-centre 3D 
synthetic spectrum (right ordinate) of the 615\pun{nm} line (solid), is compared to that 
of the 630\pun{nm} (dashed) and 777\pun{nm} (triple dot-dashed) lines. The
contribution functions are scaled to have similar maxima and are
plotted against their respective monochromatic continuum optical depths.}
\label{contf_615}
\end{figure}

Figure \ref{contf_615} shows the EW contribution function
\citep[following][]{magain86}
of the 3D disc-centre synthetic spectrum for the 615.8\pun{nm} line (solid) 
in comparison with the 630.0\pun{nm} (dashed) and 777.1\pun{nm}
(triple dot-dashed) lines, together with the temperature structures 
of the 3D and \xx\ models.
The 615.8\pun{nm} line is formed deeply inside the photosphere.

For the 1D synthetic spectra we have the possibility
of including all the relevant atomic and molecular lines and thus
perform line profile fitting.
We performed this exercise only on the Kurucz Flux:
an ATLAS+SYNTHE grid provides  
A(O)=8.75.
This result is consistent with those based on
EW measurements as reported above.

The abundance derived from this line is uncertain for the following reasons:
the continuum placement is difficult;
the atomic parameters of the blend lines in the range are not well known.
We note that \citet{hhoxy} has discarded this line in his analysis of
the solar oxygen abundance.
We decided to  keep it among possible oxygen abundance indicators,
although the large associated error gives it a low weight.


\subsubsection{\ion{O}{i} 777 nm triplet}

\begin{figure*}
\resizebox{\hsize}{!}{\includegraphics[clip=true,angle=0]{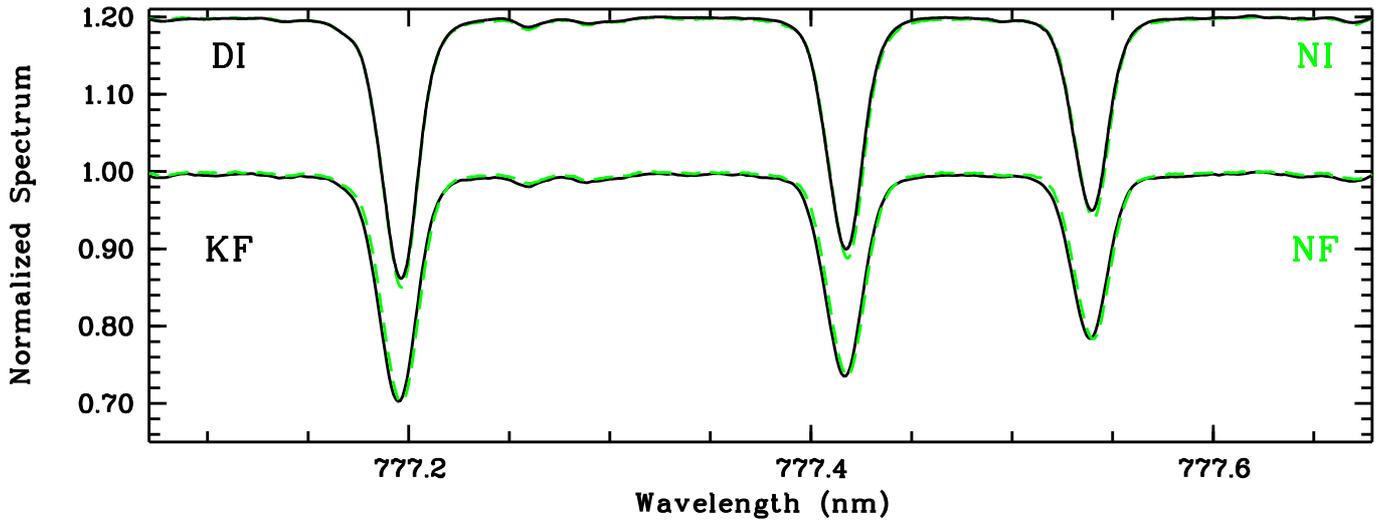}}
\caption{The \ion{O}{i} 777\pun{nm} triplet in the four observed spectra 
used in this work. Symbols are as for Fig.\,\ref{oi630}.
}
\label{oi777}
\end{figure*}

These three strong lines (see Fig.\,\ref{oi777}) lie in a clean 
region of the spectrum and do not suffer any blending
from other atomic or molecular lines, nor of any
telluric absorption. 
The strength of these lines places them on the
saturation part of the curve-of-growth which, however,
is not really flat and still provides a reasonable
abundance sensitivity for these lines.
The lines have relatively extended wings, such that
the wings of neighbouring lines overlap.
The EW of each line is affected by this effect by 
1\% to 2.5\% (the 777.4\pun{nm} line being the most affected).
The triplet lines arise from a high excitation level (9.146\pun{eV})
and are very sensitive to departures from LTE. 
From Table~\ref{oabun} it is evident that the \emph{LTE} oxygen 
abundance obtained from \cobold+\linfor\ for the three lines
is larger for the fluxes than for the intensities,
which is consistent with the fact that NLTE corrections
are larger for flux than for intensity (see Table \ref{nltecorr}).

The 3D correction (which are expected to be similar in LTE and NLTE) is 
about +0.05\,dex for flux and +0.10\,dex for intensity (assuming a
micro-turbulence $\xi$=1.0\,\kms).
ATLAS+WIDTH implies the same abundances as \cobold+\linfor, while the 
abundances using the MARCS model are 0.06\,dex {\em lower}.
Therefore we conclude, like for that of the forbidden lines,
that the direction of our 3D correction is {\em opposite}
to that found by A04
using the MARCS solar model 
as 1D reference: while we find A(3D)--A(1D$_{\rm MARCS}$)=+0.06, 
they find A(3D)--A(1D$_{\rm MARCS}$)=--0.06.

Since the lines are on the saturation part of the
curve-of-growth, they are sensitive to the
micro-turbulence $\xi$ adopted in the 1D analysis.
The difference due to a change from 1.0\,\kms to 1.5\,\kms
produces in both \mD\ and \xx\ models a difference of --0.04\,dex 
(for 777.1\pun{nm}).

The fit of the line profile is problematic.
The shape of the line is affected by
NLTE effects (A04), so even
the \cobold+\linfor\ LTE synthetic spectra can not reproduce
very well the observed shape of the line which is wider than the
synthetic spectra. 
For this reason our abundance analysis for these
lines relies on EWs and not on line profile fitting.

Our adopted LTE oxygen abundances for the three
lines of the triplet as obtained by matching the measured
EWs with the 3D curve-of-growth are the following
(see Tables~\ref{oabun} and \ref{finaboxy} for details).
From flux spectra:
\\
A(O)=$9.03\pm (0.00)_{\rm stat}\pm (0.12)_{\rm syst}$,
\\
A(O)=$8.99\pm (0.01)_{\rm stat}\pm (0.11)_{\rm syst}$, 
\\
A(O)=$8.96\pm (0.00)_{\rm stat}\pm (0.10)_{\rm syst}$.
\\
From disc-centre intensity spectra:
\\
A(O)=$8.93\pm (0.01)_{\rm stat}\pm (0.12)_{\rm syst}$, 
\\
A(O)=$8.88\pm (0.02)_{\rm stat}\pm (0.11)_{\rm syst}$, 
\\
A(O)=$8.88\pm (0.01)_{\rm stat}\pm (0.10)_{\rm syst}$.


\subsubsection{\ion{O}{i} 844 nm}

\begin{figure}
\resizebox{\hsize}{!}{\includegraphics[clip=true,angle=0]{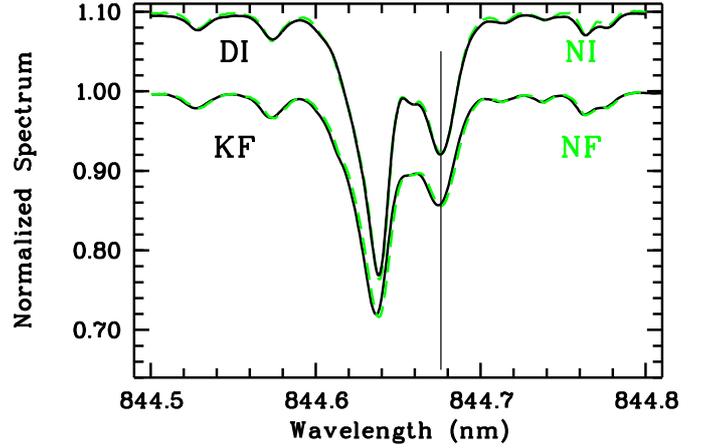}}
\caption{The \ion{O}{i} 844\pun{nm} line in the four observed spectra 
used in this work. Symbols are as for Fig.\,\ref{oi630}. The vertical line
indicated the laboratory wavelength position of the line.
}
\label{oi844}
\end{figure}

This line (see Fig.\,\ref{oi844}) is blended mainly by an iron line 
for which atomic data are not well known.
Also in this case we tried to divide
the observed spectrum by a 1D ATLAS+SYNTHE synthetic one.
The spectrum obtained, that should contain oxygen only, has a shape
which is not in agreement with the \cobold+\linfor\ synthetic
line profile in the blue side, where the \ion{Fe}{i} blend line lies,
but the agreement is good on the red side.

A de-blending EW computation is not always stable, 
however our ``de-blended'' EW provides
an oxygen abundance in the range 8.80 to 8.84,
depending on the observed spectrum considered.
The fact that an over-plot of the observed spectrum, ``as is'' with the 
\cobold+\linfor\ synthetic spectra suggests a value of 8.86 for
the oxygen abundance and that a fit with the \cobold+\linfor\
grid avoiding the blue wing gives A(O)=8.80, 
indicates that our EW is nevertheless useful for the abundance 
determination (see Table~\ref{oabun}).
The 3D correction is +0.03\,dex for flux and +0.07\,dex
for disc-centre intensity ($\xi=1.0$\,\kms).

Using a 1D ATLAS+SYNTHE grid, we fitted both the observed Kurucz Flux 
spectrum ``as is'' and the ``divided-out'' spectrum containing only 
the oxygen line (obtained by dividing by a synthetic spectrum of the blend
lines only), and obtain A(O)=8.79 and 8.73, respectively.
Using a 1D HM+SYNTHE grid in the same way, we obtain instead 
A(O)=8.82 and 8.76, respectively.

Our adopted LTE oxygen abundance from this line is:
A(O)=$8.83\pm (0.00)_{\rm stat}\pm (0.33)_{\rm syst}$ from the flux spectra, 
and
A(O)=$8.80\pm (0.02)_{\rm stat}\pm (0.33)_{\rm syst}$ from disc-centre 
intensity spectra (see Table~\ref{finaboxy} for details).


\begin{table*}
\caption{Solar LTE oxygen abundance for individual lines as derived 
from different spectra and model atmospheres.}
\label{oabun}
\begin{center}
\begin{tabular}{rrrrrrrrrrrrrr}
\noalign{\smallskip}\hline\noalign{\smallskip}
Spectrum & Wavelength & EW  &  \multicolumn{7}{c}{A(O)} &  \multicolumn{4}{c}{3Dcor}\\
 &  &  & 3D & \multicolumn{2}{c}{~~~\mD}&\multicolumn{2}{c}{~~~\xx}&\multicolumn{2}{c}{~~~HM}&\multicolumn{2}{c}{~~~3D-\mD}&
\multicolumn{2}{c}{~~~3D-\xx}\\
         & nm         & pm  &  & 1.0 & 1.5 & 1.0 & 1.5 & 1.0 & 1.5 & 1.0 & 1.5 & 1.0 &  1.5\\
(1) &    (2) &       (3) &     (4) &     (5) &    (6) &     (7) &   (8) &    (9) &   (10) &   (11) &   (12) &   (13) & (14)\\
\noalign{\smallskip}\hline\noalign{\smallskip}
 KF &  615.8 &    0.364  &   8.639 &   8.719 &  8.717 &  8.668 &  8.665 &  8.782 &  8.779 &--0.080 &--0.077 &--0.029  &--0.026\\
 NF &  615.8 &    0.365  &   8.641 &   8.721 &  8.718 &  8.669 &  8.667 &  8.783 &  8.781 &--0.080 &--0.077 &--0.028  &--0.026\\
 NI &  615.8 &    0.539  &   8.645 &   8.705 &  8.701 &  8.666 &  8.663 &  8.785 &  8.782 &--0.059 &--0.056 &--0.021  &--0.018\\
 DI &  615.8 &    0.499  &   8.606 &   8.666 &  8.663 &  8.628 &  8.626 &  8.747 &  8.745 &--0.060 &--0.057 &--0.022  &--0.020\\
\noalign{\smallskip}
 KF &  630.0 &    0.551$^1$  &   8.695 &   8.690 &  8.687 &  8.645 &  8.642 &  8.715 &  8.712 &  0.005 &  0.008 &  0.050  &  0.053\\
 NF &  630.0 &    0.530$^1$  &   8.668 &   8.663 &  8.661 &  8.615 &  8.613 &  8.688 &  8.686 &  0.005 &  0.007 &  0.053  &  0.055\\
 NI &  630.0 &    0.470$^1$  &   8.686 &   8.667 &  8.665 &  8.613 &  8.611 &  8.698 &  8.696 &  0.019 &  0.021 &  0.073  &  0.075\\
 DI &  630.0 &    0.457$^1$  &   8.666 &   8.646 &  8.644 &  8.591 &  8.589 &  8.678 &  8.677 &  0.020 &  0.022 &  0.075  &  0.077\\
\noalign{\smallskip}
 KF &  636.4 &    0.146  &   8.780 &   8.772 &  8.772 &  8.756 &  8.756 &  8.786 &  8.786 &  0.008 &  0.008 &  0.024  &  0.024\\
 NF &  636.4 &    0.153  &   8.801 &   8.793 &  8.793 &  8.777 &  8.777 &  8.807 &  8.806 &  0.008 &  0.008 &  0.024  &  0.024\\
 NI &  636.4 &    0.139  &   8.837 &   8.822 &  8.822 &  8.802 &  8.802 &  8.839 &  8.839 &  0.015 &  0.015 &  0.035  &  0.035\\
 DI &  636.4 &    0.106  &   8.718 &   8.703 &  8.703 &  8.683 &  8.683 &  8.721 &  8.720 &  0.015 &  0.015 &  0.035  &  0.035\\
\noalign{\smallskip}
 KF &  777.1 &    8.140  &   9.034 &   9.025 &  8.984 &  8.974 &  8.935 &  9.030 &  8.988 &  0.009 &  0.050 &   0.059 &  0.099\\
 NF &  777.1 &    8.080  &   9.026 &   9.017 &  8.976 &  8.966 &  8.927 &  9.022 &  8.980 &  0.009 &  0.050 &   0.059 &  0.099\\
 NI &  777.1 &    9.290  &   8.952 &   8.912 &  8.870 &  8.841 &  8.802 &  8.934 &  8.892 &  0.040 &  0.081 &   0.111 &  0.150\\
 DI &  777.1 &    8.900  &   8.903 &   8.863 &  8.824 &  8.794 &  8.755 &  8.886 &  8.846 &  0.040 &  0.080 &   0.110 &  0.149\\
\noalign{\smallskip}
 KF &  777.4 &    6.870  &   8.992 &   8.987 &  8.948 &  8.939 &  8.902 &  8.992 &  8.953 &  0.005 &  0.044 &   0.053 &  0.091\\
 NF &  777.4 &    6.800  &   8.981 &   8.976 &  8.937 &  8.928 &  8.891 &  8.981 &  8.942 &  0.005 &  0.044 &   0.053 &  0.090\\
 NI &  777.4 &    7.870  &   8.912 &   8.875 &  8.836 &  8.807 &  8.770 &  8.899 &  8.860 &  0.037 &  0.076 &   0.105 &  0.142\\
 DI &  777.4 &    7.350  &   8.838 &   8.802 &  8.765 &  8.735 &  8.698 &  8.827 &  8.791 &  0.035 &  0.074 &   0.103 &  0.140\\
\noalign{\smallskip}
 KF &  777.5 &    5.420  &   8.967 &   8.967 &  8.932 &  8.922 &  8.889 &  8.974 &  8.940 &  0.000 &  0.035 &   0.045 &  0.078\\
 NF &  777.5 &    5.340  &   8.953 &   8.953 &  8.918 &  8.908 &  8.875 &  8.960 &  8.926 &  0.000 &  0.034 &   0.045 &  0.078\\
 NI &  777.5 &    6.280  &   8.895 &   8.863 &  8.829 &  8.800 &  8.766 &  8.891 &  8.857 &  0.032 &  0.066 &   0.095 &  0.129\\
 DI &  777.5 &    6.090  &   8.864 &   8.833 &  8.800 &  8.769 &  8.737 &  8.861 &  8.830 &  0.030 &  0.065 &   0.094 &  0.127\\
\noalign{\smallskip}
 KF &  844.6 &    3.511  &   8.830 &   8.848 &  8.823 &  8.803 &  8.779 &  8.864 &  8.841 &--0.017 &  0.007 &  0.027  &  0.051\\
 NF &  844.6 &    3.533  &   8.836 &   8.853 &  8.828 &  8.808 &  8.784 &  8.869 &  8.846 &--0.017 &  0.007 &  0.028  &  0.052\\
 NI &  844.6 &    4.372  &   8.789 &   8.777 &  8.752 &  8.716 &  8.692 &  8.820 &  8.796 &  0.011 &  0.037 &  0.072  &  0.096\\
 DI &  844.6 &    4.492  &   8.813 &   8.801 &  8.775 &  8.739 &  8.715 &  8.842 &  8.818 &  0.012 &  0.038 &  0.074  &  0.098\\
\noalign{\smallskip}
 WI &  926.6 &    3.829  &   8.736 &   8.744 &  8.731 &  8.677 &  8.664 &  8.799 &  8.787 & -0.008 &  0.005 &  0.060  &  0.072\\
\noalign{\smallskip}
 AI & 1130.2 &    1.920  &   8.748 &   8.759 &  8.752 &  8.706 &  8.700 &  8.824 &  8.818 &--0.011 &--0.004 &  0.042  &  0.048\\
\noalign{\smallskip}
 AI & 1316.5 &    1.980$^2$ &   8.767 &   8.767 &  8.762 &  8.740 &  8.735 &  8.835 &  8.830 &  0.000 &  0.005 &  0.027  &  0.031\\
 AI & 1316.5 &    3.150$^3$ &   8.763 &   8.766 &  8.763 &  8.741 &  8.739 &  8.833 &  8.830 & -0.003 &  0.000 &  0.022  &  0.024\\
\noalign{\smallskip}\hline\noalign{\smallskip}
\end{tabular}
\end{center}
Col.~(1) spectrum identification: DI: Delbouille Intensity, 
NI: Neckel Intensity, WI: Wallace Intensity, AI: Ayres Intensity, 
NF: Neckel Flux, KF: Kurucz Flux.
Col.~(2) wavelength of the line.
Col.~(3) measured equivalent width.
Col.~(4) oxygen abundance, A(O), according to the \cobold\ 3D model.
Cols.~(5)--(10) A(O) from 1D models, odd numbered cols.\ correspond
to a micro-turbulence~$\xi$ of 1.0\,\kms, even numbered cols.\ to 
$\xi = 1.5$\,\kms.
Col.~(11)-(14) 3D corrections, even numbered cols.\ for
$\xi = 1.0$\,\kms, and odd numbered cols.\ for 1.5\,\kms, respectively.
\\
(1) EW of the blend [\ion{O}{i}]+\ion{Ni}{i}; the contribution to the EWs due
to Ni for A(Ni)=6.25 is 0.20 and 0.18\pun{pm}
for flux and intensity spectra, respectively, according to our 3D investigation.
(2) EW relative to the two blending \ion{O}{i} lines of the multiplet.
(3) EW relative to the complete multiplet.
\end{table*}

\begin{table*}
\caption{Oxygen LTE abundance for each line with the error estimates. The
second and third column are the LTE abundance using the \cobold\ 3D and HM 
model, respectively. The last column indicates whether a flux (F) or
intensity (I) measurement was used.}
\label{finaboxy}
\begin{center}
\begin{tabular}{rcccccccc}
\noalign{\smallskip}\hline\noalign{\smallskip}
 line   & A(O)$_{\rm 3D}$ & A(O)$_{\rm HM}$      & statistical & \loggf & broadening & blend  & total & Spectrum \\
\pun{nm}&      &      & error     & error       & error  & error      & error  &  \\
\noalign{\smallskip}\hline\noalign{\smallskip}
 615.8  & 8.64 & 8.78 & 0.01      & 0.16   &  0.01      &  0.14  &  0.21 & F   \\
 615.8  & 8.63 & 8.77 & 0.02      & 0.16   &  0.01      &  0.14  &  0.21 & I   \\
 630.0  & 8.68 & 8.69 & 0.02      & 0.03   &  0.00      &  0.15  &  0.15 & F/I \\
 636.3  & 8.78 & 8.79 & 0.05      & 0.03   &  0.00      &  0.12  &  0.12 & F/I \\
 777.1  & 9.03 & 9.03 & 0.00      & 0.07   &  0.06      &  0.08  &  0.12 & F    \\
 777.1  & 8.93 & 8.91 & 0.01      & 0.07   &  0.07      &  0.05  &  0.11 & I    \\
 777.4  & 8.99 & 8.99 & 0.01      & 0.07   &  0.05      &  0.06  &  0.11 & F   \\
 777.4  & 8.88 & 8.86 & 0.02      & 0.07   &  0.06      &  0.06  &  0.11 & I   \\
 777.5  & 8.96 & 8.97 & 0.00      & 0.07   &  0.04      &  0.06  &  0.10 & F   \\
 777.5  & 8.88 & 8.88 & 0.01      & 0.07   &  0.03      &  0.06  &  0.10 & I   \\
 844.6  & 8.83 & 8.87 & 0.00      & 0.23   &  0.03      &  0.24  &  0.33 & F   \\
 844.6  & 8.80 & 8.83 & 0.00      & 0.23   &  0.03      &  0.24  &  0.33 & I   \\
 926.6  & 8.74 & 8.80 & 0.01      & 0.07   &  0.02      &  0.09  &  0.12 & I   \\
1130.2  & 8.75 & 8.82 & 0.01      & 0.07   &  0.01      &  0.15  &  0.17 & I    \\
1316.4  & 8.76 & 8.82 & 0.01      & 0.16   &  0.01      &  0.15  &  0.22 & I   \\
\noalign{\smallskip}\hline\noalign{\smallskip}
\end{tabular}
\end{center}
\end{table*}

\begin{table*}
\caption{NLTE corrections for all the lines,
for different values of ${\rm S_H}$. The corrections
from \citet{hhoxy} and  A04 are also
given for comparison.}
\label{nltecorr}
\begin{center}
\begin{tabular}{rcclcclclcc}
\noalign{\smallskip}\hline\noalign{\smallskip}
 line  & LTE &\multicolumn{3}{c}{NLTE(${\rm S_H}=1$)} &\multicolumn{2}{c}{NLTE(${\rm S_H}=1/3$)}&
\multicolumn{3}{c}{NLTE(${\rm S_H}=0$)}   & Flux/Intensity \\
\pun{nm} & A(O) & A(O) & corr. & corr. HH$^a$ & A(O) & corr. & A(O) & corr. & corr. A04$^b$\\
\noalign{\smallskip}\hline\noalign{\smallskip}
     615.8 & 8.64 &  8.64 &--0.003 &        & 8.64 &--0.004&  8.64 & --0.002&--0.03 & F\\
     615.8 & 8.63 &  8.62 &--0.003 &--0.01  & 8.62 &--0.003&  8.62 & --0.002&       & I\\
     630.0 & 8.68 &  8.68 &  ~~0.0 &        & 8.68 &~~0.0  &  8.68 & ~~0.0  &       &\\
     636.3 & 8.78 &  8.78 &  ~~0.0 &        & 8.78 &~~0.0  &  8.78 & ~~0.0  &       &\\
     777.1 & 9.03 &  8.87 &--0.16  &        & 8.81 &--0.22 &  8.75 & --0.28 &--0.27 & F\\
     777.1 & 8.93 &  8.85 &--0.07  &--0.07  & 8.81 &--0.12 &  8.77 & --0.16 &       & I\\
     777.4 & 8.99 &  8.84 &--0.14  &        & 8.79 &--0.20 &  8.74 & --0.25 &--0.24 & F\\
     777.4 & 8.88 &  8.80 &--0.08  &--0.06  & 8.76 &--0.12 &  8.72 & --0.16 &       & I\\
     777.5 & 8.96 &  8.85 &--0.12  &        & 8.80 &--0.16 &  8.75 & --0.21 &--0.20 & F\\
     777.5 & 8.88 &  8.82 &--0.06  &--0.05  & 8.79 &--0.09 &  8.75 & --0.13 &       & I\\
     844.6 & 8.83 &  8.75 &--0.08  &        & 8.72 &--0.12 &  8.68 & --0.15 &--0.20 & F\\
     844.6 & 8.80 &  8.75 &--0.05  &        & 8.72 &--0.08 &  8.70 & --0.11 &       & I\\
     926.6 & 8.74 &  8.71 &--0.03  &--0.02  & 8.69 &--0.05 &  8.65 & --0.08 &--0.08 & I\\
    1130.2 & 8.75 &  8.73 &--0.01  &--0.01  & 8.72 &--0.02 &  8.71 & --0.04 &       & I\\
    1316.4 & 8.76 &  8.75 &--0.01  &--0.01  & 8.75 &--0.01 &  8.74 & --0.02 &       & I\\
\noalign{\smallskip}\hline\noalign{\smallskip}
\end{tabular}
\\
$^a$ 1D-NLTE correction from \citet{hhoxy}\\
$^b$ 3D-NLTE correction from A04
\end{center}
\end{table*}


\subsubsection{\ion{O}{i} 926 nm}

\begin{figure}
\resizebox{\hsize}{!}{\includegraphics[clip=true,angle=0]{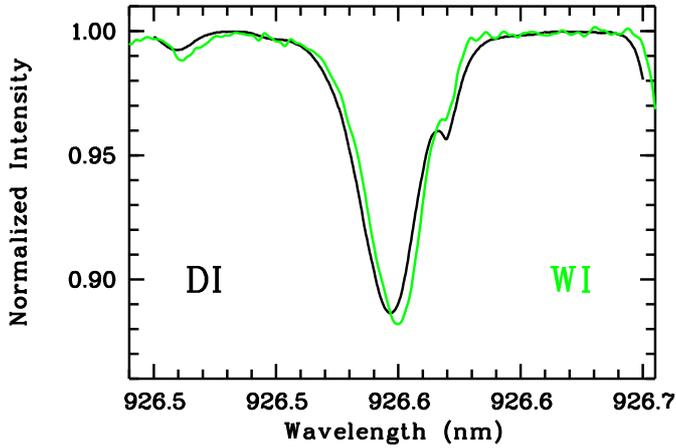}}
\caption{The observed Delbouille (DI) and Wallace (WI) 
disc-centre intensity spectra of the \ion{O}{i} 926\pun{nm} line.
}
\label{oi926}
\end{figure}

In the Kurucz Flux spectrum, as in the Neckel Flux and Intensity spectra, 
the \ion{O}{i} 926 nm line is contaminated by a telluric line.
The intensity spectra of Delbouille (see Fig.\,\ref{oi926}) and Wallace
have been processed to subtract the telluric contribution.
The  subtraction is probably not completely
successful, since the profile still shows some distortion.
Nevertheless, the spectra can be fitted reasonably well. Although the 
Delbouille spectrum has a higher signal-to-noise (about 5000), the 
Wallace spectrum (S/N about 500) shows less residual distortion after
the subtraction of the telluric line. For this reason 
the EW was computed only from the Wallace intensity spectrum,
providing an oxygen abundance of A(O)=8.75 in 3D.
The 3D correction in this case is +0.06\,dex ($\xi=1.0$\,\kms).

We also did line fitting restricted to a range that excludes the 
portion of the line that is affected by the telluric absorption.
The fit of the Delbouille spectrum with a \cobold+\linfor\ grid
gives A(O)=8.67. The 3D fitting of the Wallace spectrum results in
A(O)=8.71. Fitting the Delbouille and Wallace spectra with an 
ATLAS+SYNTHE grid, we obtain A(O)=8.68 and 8.66, respectively.
With the Holweger-M\"uller+SYNTHE grid, we have instead A(O)=8.74 
and 8.73, respectively\footnote{Note, that the values given in Tab.~\ref{oabun}
differ from the values here since they are based on matching the line EW, not
line profile fitting.}.

For this line our adopted LTE oxygen abundance is
A(O)=$8.74\pm (0.01)_{\rm stat}\pm (0.12)_{\rm syst}$
(see Table~\ref{finaboxy} for details). 


\subsubsection{\ion{O}{i} 1130 nm}
\label{OI1130}

\begin{figure}
\resizebox{\hsize}{!}{\includegraphics[clip=true,angle=0]{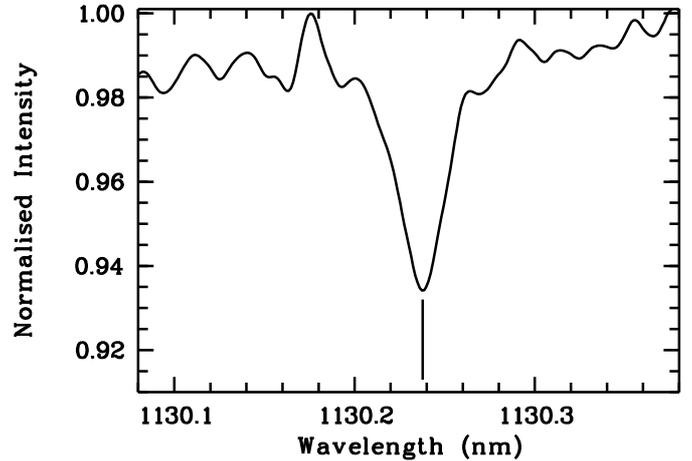}}
\caption{The observed disc-centre intensity spectrum we have used for the 
\ion{O}{i} 1130\pun{nm} line. 
The vertical line indicates  the laboratory wavelength position of the line.
}
\label{oi1130}
\end{figure}

In the Kurucz Flux atlas and in both Neckel spectra, 
the \ion{O}{i} 1130\pun{nm} line is blended by a telluric 
absorption. The Wallace intensity spectrum has been cleaned from
telluric absorption, but its S/N is only about 80, 
inferior to the Kitt Peak FTS scans we considered, and insufficient
for our purposes.

Our analysis of this line relies therefore on a special reduction of FTS
scans obtained from the NSO Digital Archive (see Fig.\,\ref{oi1130}).  
The \ion{O}{i} 1130\pun{nm} line, and the companion 1316\pun{nm} multiplet 
described below, fall in the shoulders of very strong telluric absorptions, 
and require a careful correction for the blending. 
One of us (TRA) examined the NSO/FTS catalogue of 
observations and identified a suitable series of scans, taken on the 13th
and 14th of September 1983.  Each scan was a co-add of two interferograms
totalling 5 minutes exposure, through a 5$^{\prime\prime}$ diameter
aperture, at disc-centre. The datasets are listed as ``8309NNr0.XXX'' 
where ``NN'' is the day (13 or 14), and ``XXX'' is a scan number.  Hereafter
these will be described as ``NN.XXX' for simplicity.  Scan 13.004 was
taken in the early morning at about 09:00 local time, at a moderate
airmass of 1.7. Scan 13.011 was acquired close to local noon at a low
airmass of 1.14. Scan 13.015 was taken later in the afternoon, again
at a moderate airmass of 1.5.  On the following day, scan 14.007 was
collected at about 09:00 local time at an airmass of 1.7, while
scan 14.010 was taken about an hour before local noon, at a lower airmass 
of 1.27.  The spectral resolution of these data is very high, around 
$2{\times}10^{5}$, and the S/N is very high as well, $>10^{3}$,
taking into account only the Poisson noise and ignoring other
sources of uncertainty. 
Owing to the excellent frequency stability of the FTS scans, pairs taken at 
low and high airmass can be registered very precisely to the telluric
absorptions, and ratioed to eliminate the solar component. The ratio
can be modelled as an exponential attenuation factor, and then applied
back to the observed spectra as a correction for the telluric component.
By subjecting this technique to various combinations of the low and high
airmass scans, one can average over slight variations of the atmospheric
attenuation factor in the morning versus the afternoon (owing, say, to
increases in the humidity due to solar heating), and possible slight
changes in the solar component due to 5 minute oscillations, convection,
and the like. 
All these effects contribute to the deterioration of the information
content of the spectrum and may be effectively thought of, as ``noise''.
This source of noise
is in fact much larger than the pure Poisson noise in the spectrum.
The dispersion in the derived solar profiles 
is a rough measure of the uncertainty of the telluric correction
procedure, and of intrinsic solar variations. For the \ion{O}{i} 1130\pun{nm}
line, the corresponding S/N is about 15.
We refer to this S/N ratio as ``equivalent S/N''.

The line is blended with a weaker iron line at $\lambda =$1130.242 nm, 
\loggf=$-2.381$ from Kurucz's Cowan-code computations, 
whose contribution is 17\% (for a ATLAS+SYNTHE synthetic spectrum
with A(O)=8.60) to the complete EW of the blend.

From fitting with a \cobold+\linfor\ grid we derive an oxygen abundance 
of A(O)=8.76. The corresponding value obtained from fitting with ATLAS+SYNTHE
is A(O)=8.75. The EW based abundance determination gives a fully consistent
result, A(O)=8.75. The 3D correction is +0.04\,dex ($\xi=1.0$\,\kms).

Our adopted LTE oxygen abundance for this line is:
A(O)=$8.75\pm (0.01)_{\rm stat}\pm (0.17)_{\rm syst}$
(see Table~\ref{finaboxy} for details).


\subsubsection{\ion{O}{i} 1316 nm}

\begin{figure}
\resizebox{\hsize}{!}{\includegraphics[clip=true,angle=0]{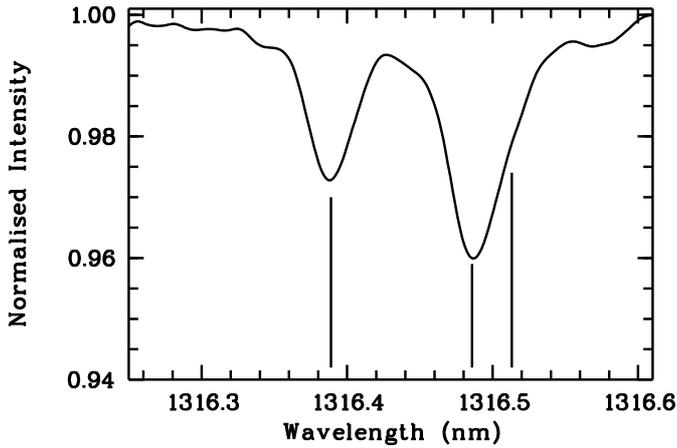}}
\caption{The observed disc-centre intensity spectrum we have used for the 
analysis of the \ion{O}{i} 1316\pun{nm} line. The vertical lines indicate
the laboratory wavelengths of the three \ion{O}{i} lines discussed in the text.
}
\label{oi1316}
\end{figure}

As shown in Fig.\,\ref{oi1316}, three oxygen lines can be found in this Multiplet,
at $\lambda$ 1316.3889\pun{nm}, 1316.4858\pun{nm}, and 1316.5131\pun{nm}.
The strongest one (1316.4858\pun{nm}) is blended
with one of the other oxygen lines (1316.5131\pun{nm}).
For the three lines we adopt for \loggf~ values of 
$-0.28$, $-0.250$, $-0.728$, respectively, according to \citet{biemont91}.

The Wallace intensity solar spectrum
has a S/N ratio of about 700, but the spectrum is poorly rectified,
making the continuum placement very uncertain. We decided not to use
this spectrum but rather the Kitt Peak FTS scans mentioned above 
(Sect.\,\ref{OI1130}), 
which have a higher S/N ratio ($> 10^3$) and are well rectified.
For the 1316\pun{nm} feature the equivalent 
S/N for the strongest of the lines, having a central depth of about 5\%, 
is about 25, somewhat higher than for the 1130\pun{nm} line.
From fitting this spectrum,
we obtain A(O)=8.75 with a 3D grid, for all the three lines.
With an ATLAS+SYNTHE grid, the fit gives A(O)=8.77.

For this line our LTE adopted abundance is 
A(O)=$8.76\pm (0.01)_{\rm stat}\pm (0.22)_{\rm SST}$.


\section{Discussion and conclusions}
\label{disc_concl}

The best estimate of the solar oxygen abundance 
is obtained by a suitable average of the abundances
derived from the individual lines.
Our adopted oxygen abundances are summarised in
Table~\ref{finaboxy} together with associated
errors, while corrections for departures
from LTE are listed in Table~\ref{nltecorr}. 
A different choice in the collision efficiency with H-atoms,
quantified by the scaling factor ${\rm S_H}$ (see Sec.5.1), 
produces a different value in the solar oxygen abundance.
A weighted mean 

\begin{equation}
\begin{array}{c @{~~~~~~~~~~~~~~~~~~~~~~} c}
\mu = {\Sigma\left(x_i/\sigma_i^2\right)\over\Sigma\left(1/\sigma_i^2\right)}
&
\sigma^2_\mu = {1\over\Sigma\left(1/\sigma_i^2\right)}
\end{array}
\end{equation}

\noindent
\citep[][p. 59]{bevington},
using the total errors from Table~\ref{finaboxy},
provides an oxygen abundance of 

\begin{equation}
\begin{array}{c @{\rm A(O)=~} c @{~~~~~{\rm for} ~~~~~ {\rm S_H} =} c }
     & 8.73\pm 0.04 & 0\\
     & 8.76\pm 0.04 & 1/3\\
     & 8.79\pm 0.04 & 1\\
\end{array}
\end{equation}

\noindent
Our preference is an intermediate value, $S_H$=1/3,
so that the solar oxygen abundance we advocate is 
A(O)=$8.76\pm 0.04\pm 0.03$.
The first error includes the statistical uncertainties as well as 
any systematic uncertainty, except the one related to the NLTE
corrections, which are considered separately to be 0.03\,dex.
If we sum the two errors, we arrive at an oxygen
abundance of \boldmath {\bf A(O)=$8.76\pm 0.07$}. \unboldmath
Computing A(O) from the simple average, instead of taking the 
weighted mean, we obtain only slightly smaller abundances.
Oxygen abundances derived from intensity spectra are systematically 
smaller than those obtained from flux spectra, but the difference is 
less than 0.04\,dex.

If we consider the NLTE abundance in column~6 of Table~\ref{nltecorr},
corresponding to ${\rm S_H}=1/3$, excluding the 615\pun{nm} line, we find a
line to line scatter of 0.13\,dex, not incompatible with the 1$\sigma$ error
bars we attribute to the individual lines.  The same range in A(O) is
obtained for ${\rm S_H}=0$.  The 615\pun{nm} line requires a lower abundance
and would increase the overall scatter, but, as we discussed in the section
devoted to this line, this line is particularly problematic.

The lowest abundance after the 615\pun{nm} line is coming from the forbidden
oxygen line at 630\pun{nm}. The relatively low oxygen abundance it implies
is sensitively dependent on the abundance of nickel which we {\em assumed\/}
to be 6.25, the presently recommended value.  Had be adopted the equivalent
width of 4.3\pun{pm} which A04 attribute to the oxygen contribution to the
630\pun{nm} line we would have obtained A(O)=8.78 -- moving the result closer to the
other lines and becoming fully consistent with forbidden oxygen line at 636\pun{nm}.

We do not show comparisons between line profile shapes of synthetic and observed spectra
because we cannot reproduce the observed features due to blends, as
is the case for the 636, 844, and 926\pun{nm} lines; the triplet is
affected by NLTE corrections. Since our NLTE computation is based on the \mD\
model, we cannot reproduce the profile of the triplet lines accurately. The
infra-red lines, after correction for telluric absorption, have low S/N so
that visual comparison to synthetic profiles is not particularly illuminating. Only
for the 630\pun{nm} forbidden line the observed line profile shape can be
compared to the synthetic 3D spectrum, and this is shown in
Fig.~\ref{fit3d_o630} for all four observed profiles.

Our results must be compared with those
of \citet{hhoxy}, A04, and to the one
of \citet{ayres08}.
\citet{hhoxy} provides A(O)$=8.736\pm 0.078$,
thus $\sim$0.05\,dex lower than our weighted
mean for ${\rm S_H}=1$, right within one $\sigma$ of his error bar.
The mean oxygen abundance obtained
from the atomic lines  by A04, 
A(O)=8.65, and our weighted mean, for ${\rm S_H}=0$, differ of +0.08\,dex.

One of us \citep{ayres08} has recently carried out an independent analysis of
specifically the 630\pun{nm} \ion{O}{i} forbidden line, taking a different
approach than followed in the present work. Here, we have allowed the
wavelength scale in the vicinity of the 630\pun{nm} feature to float, to
accommodate the \ion{Ni}{i} blend at its nominal abundance. With this approach
we obtain an excellent match to the observed line-shape, although \ion{Ni}{i}
accounts for a larger fraction of the equivalent width of the full feature
(see Fig.\,\ref{o630blend}) than would be inferred from the original analysis
of the forbidden oxygen line by \citet[][their Fig.\,1]{ALA01}. Curiously, we
obtain nearly the same oxygen abundance as \citet{ALA01}, even though here the
\ion{O}{i} component occupies a smaller fraction of the full equivalent width.
At the time \citet{ALA01} performed their analysis the oscillator strength of
the Ni transition was not precisely known, and they left the product
$gf\epsilon(\mathrm{Ni})$ as a free fitting parameter. They obtained a
contribution corresponding to a low nickel abundance of A(Ni)=6.05, i.e. a
rather small contribution, using todays accepted value of the oscillator
strength of the Ni transition. It is perceivable that the difficulty in the
continuum placement are responsible for the disparate findings.

\citet{ayres08} took the alternative fitting
approach of fixing the wavelength scale and allowing the \ion{Ni}{i} abundance
to float.  Although the absolute velocity calibrations of the solar atlases and
specially reduced spectra we have described earlier can be uncertain
by hundreds of m~s$^{-1}$, Ayres circumvented that deficiency by adjusting
an observed FTS spectrum to match 3D-synthesised cores of several \ion{Fe}{i}
absorption lines in the immediate vicinity of the [\ion{O}{i}] 630\pun{nm} feature 
(the 3D model was a single CO$^5$BOLD snapshot in this case). This procedure 
moves the blend blue-ward of the result we obtained here, leading to a higher 
fraction of the equivalent width contributed by \ion{O}{i}, and lowering the 
fraction of \ion{Ni}{i} (by about 30\% ). The inferred oxygen abundance
is larger than obtained here, A(O)=$8.81{\pm}0.04$.  Ayres's approach relies
on an accurate prediction of the relative core velocities of the \ion{Fe}{i}
and \ion{O}{i} (+ \ion{Ni}{i}) features by the 3D model, and assumes that the 
\ion{Ni}{i} abundance can be treated as a free parameter. 

In both cases, it is
important to note that the \cobold\ model provides a different 
conversion between fractional \ion{O}{i} equivalent width and inferred oxygen
abundance than the 3D simulations reported by \citet{ALA01}
and subsequent studies also using the Stein-Nordlund class of models.  
The differences might be related to the different temperature gradients
of the 3D models (e.g., as exhibited by the 1D average stratifications shown
in Fig.\,\ref{f:coboldasplund}) since a main factor governing the strength 
of a weak photospheric line is the temperature gradient in the line 
formation zone. Another factor introducing systematic uncertainties is
the spectral synthesis. \citet{melendez04}
applied the spectral synthesis code MOOG \citep{sneden1973} in his determination
of the solar oxygen abundance. Besides mainly targeting OH molecular lines in
the infrared, he recalculated the oxygen abundance from the forbidden oxygen
line at 630\pun{nm} based on the average 3D model of A04 and their
equivalent width as well as oscillator strength obtaining A(O)=8.75. According to our calculations
horizontal temperature fluctuations have essentially no effect on the resulting
equivalent width of the line in disc integrated spectra so that Mel{\'e}ndez' value can be directly compared to the
value of 8.69 quoted by A04. We are left with the uncomfortable conclusion
that differences in the spectral synthesis can introduce differences as
large as 0.06\pun{dex} in the derived oxygen abundance.

The weighted mean of the abundances derived from several
\ion{O}{i} lines in the solar spectrum implies a solar
oxygen abundance in the range {\bf A(O)=8.73 to A(O)=8.79,} depending
on the actual influence of collisions by neutral hydrogen atoms
on the NLTE level populations of oxygen. The lower abundance is 
obtained by a total neglect of such collisions, the higher 
abundance is obtained by assuming the rather large cross sections
provided by the recipe of \citet{SH}, which may be considered 
an upper limit to the efficiency of these collisions.
Any treatment of neutral hydrogen collisions intermediate between 
these two extremes will provide an oxygen abundance in the above range.

Apart from this systematic uncertainty of the NLTE corrections, we 
regard the determination of A(O) as very robust, with a total error in the
weighted mean of 0.07\,dex. Considering our present ignorance on the 
actual role of collisions by neutral hydrogen, we believe that it is 
reasonable to assume that the true value lies somewhere in between these 
two extreme assumptions.
Therefore, assuming ${\rm S_H}\approx$1/3, our recommended value
for the solar oxygen abundance is the mean of the two, 
\boldmath \mbox{\bf A(O) $=8.76\pm 0.07$}. \unboldmath

Our recommended oxygen abundance is a downward revision
with respect to that of \citet{gs}. Using that set of
solar abundances the solar metallicity is $Z_\odot = 0.017$.
If we take that set of solar abundances and change only the oxygen
abundance to our recommended value, $Z_\odot$ drops to 0.016.
On the other hand our oxygen abundance is an upward revision
relative to  what was recommended by 
A04. By taking 
the solar abundances of \citet{sunabboasp} but our oxygen abundance,
we find $Z_\odot = 0.014$. Clearly the latter result is influenced 
by the general downward revision of all the solar 
abundances advocated by Asplund and collaborators. 
Having found a higher solar oxygen abundance by using \cobold\ models,
and having been unable to confirm the downward revision of phosphorus 
\citep{Ppaper} and sulphur \citep{zolfito,zolfetto},
it seems not unlikely that also for other elements a \cobold\ based
analysis would imply higher abundances than recommended by \citet{sunabboasp}. 

We should compare our result with solar oxygen abundances deduced
by other means. \citet{delahaye06} obtain A(O)=$8.86\pm 0.041\pm 0.025$(CNNe)
from the seismic properties; the latter uncertainty is retalted to the abundance of C, N, and Ne.
\citet{antia} using the helioseismically measured sound speed profile obtained $Z=0.0172\pm 0.002$,
but the method does not permit to deduce the abundance of a single
element.
\citet{chaplin07} using the frequencies of low-degree acoustic oscillations of the Sun
obtained $Z$ in the range 0.0187 - 0.0229.
\citet{basu} find
$Z$\,=\,0.016, again from helioseismic considerations.
As we can see all the results obtained from the properties of the solar
interior favour a ``high'' solar metallicity, and therefore a ``high''
oxygen abundance.
In the same line are the results obtained in the
work of \citet{stc04}: they find a better agreement
when they compare the seismic model (Z=0.017) to the solar model with
the abundances of \citet{hhoxy} (Z=0.015) with respect to the model
with A04 abundances (Z=0.012).

The downward revision of the solar $Z$ by \citet{sunabboasp}
is generally considered to be mainly due to the use of
3D models, as opposed to 1D models used in older 
analysis \citep[e.g.][]{carlos07}. 
The result of our analysis reveals, however, that at least for oxygen
this is not the case. The 3D corrections for the \ion{O}{i} lines are small
and for all lines, except for the 615.8\pun{nm} line, they are {\em positive},
indicating that the 3D model provides a {\em higher} abundance
than the corresponding \xx\ model (see Table~\ref{oabun}).
The average 3D correction, if we consider all the oxygen lines,
is $+0.049$ assuming a micro-turbulence $\xi =1.0$\,\kms in the \xx\ model, 
and $+0.066$ when $\xi =1.5$\,\kms.

\balance
Closer inspection shows that the lower oxygen abundance of A04
is essentially a result of using, {\bf on average}, lower equivalent widths and
of the choice of ${\rm S_H}=0$ (larger NLTE corrections), which are both
acting in the downward direction. The differences between the Stein \&
Nordlund model employed by A04
and our \cobold\ model (see
Fig.\,\ref{f:coboldasplund}) are not essential for the resulting oxygen
abundance. The fact that our oxygen abundance with ${\rm S_H}=0$ is 0.08 dex
higher is entirely due to our measured equivalent widths which are larger.  If
we use the \cobold\ model and the equivalent widths of A04 
we find A(O)=8.67 when averaging the oxygen abundances for the atomic lines, very
close to their result.  Therefore, the 3D corrections quoted by
A04 are in the opposite direction with respect to ours due to
the different choice of 1D reference model.

A04 argued that their suggested low O abundance would place the
Sun on the same oxygen level as is observed in the local interstellar medium
and near-by B~stars. However, it is not at all obvious that the Sun's chemical
composition should reflect its immediate surroundings since it might have
migrated towards larger galactic radii from its place of formation
\citep{WFD96} of higher metallicity. Moreover, the Sun harbours a planetary
system and it is now well established that planet-hosting stars have, on
average, higher metallicities than stars with no planets.  \citet{ecuvillon}
presented a uniform analysis of oxygen abundances in a volume-limited sample
of 59 stars with no planets and compared it to a sample of 96 planet-hosting
stars.  The mean oxygen abundance of the sample of planet-hosting stars is
8.77, while the mean oxygen abundance of the sample of stars with no planets
is 8.67.  Thus our recommended solar oxygen abundance, places the Sun in a
very popular position among planet-hosting stars in the solar neighbourhood.

To improve our knowledge of the solar oxygen abundance, one should focus in
the first place on the collisions with hydrogen atoms, both theoretically and
experimentally.  To the readers who have been following for some time the saga
of the solar abundances it is perhaps not unsurprising to find that the
measurement of high accuracy equivalent widths is still an open issue.  One
should try to better constrain the placement of the continuum, and the
influence of the blending lines.


\begin{acknowledgements}
This paper is dedicated to the late Hartmut Holweger who accompanied the
initial phases of the oxygen project with encouragement and advice.  We
thank Sveneric Johansson for his input on the isotopic splitting of nickel
lines.  We are grateful to Martin Asplund for providing us with full details of
his analysis.  Furthermore, we are grateful
to Robert Kurucz for providing the revised version of his solar flux atlas.
We wish to thank Rosanna Faraggiana for many fruitful discussion on the
topic of abundance determinations. Special thanks also to Inga Kamp for 
giving us the Kiel code and the oxygen model atom and
helping us in their use.
EC, HGL and PB acknowledge support from
EU contract MEXT-CT-2004-014265 (CIFIST).
\end{acknowledgements}


\bibliographystyle{aa}

\end{document}